\documentclass[10pt]{iopart}
\usepackage{graphicx}
\usepackage{graphicx}% Include figure files
\usepackage{epsf}
\def\agt{\mathrel{\raise.3ex\hbox{$>$}\mkern-14mu\lower0.6ex\hbox{$\sim$}}}
\def\alt{\mathrel{\raise.3ex\hbox{$<$}\mkern-14mu\lower0.6ex\hbox{$\sim$}}}

\newcommand{\beq}{\begin{equation}}
\newcommand{\eeq}{\end{equation}}
\newcommand{\beqn}{\begin{eqnarray}}
\newcommand{\eeqn}{\end{eqnarray}}

\begin{document}

\title[Collapse of magnetized, 
differentially rotating neutron stars]{Collapse and black hole 
formation in magnetized, differentially rotating neutron stars}

\author{B C Stephens$^1$, M D Duez$^1$\footnote{Present 
address:  Center for Radiophysics and Space Research,
Cornell University, Ithaca, NY 14853.}, Y T Liu$^1$, 
S L Shapiro$^1$\footnote{Also at the Department of 
Astronomy and NCSA, University of Illinois at 
Urbana-Champaign, Urbana, IL 61801.} and M Shibata$^2$}

\address{$^1$ Department of Physics, University of Illinois at 
Urbana-Champaign, Urbana, IL 61801, USA}
\address{$^2$ Graduate School of Arts and Sciences, 
University of Tokyo, Komaba, Meguro, Tokyo 153-8902, Japan}

\begin{abstract}
The capacity to model magnetohydrodynamical (MHD) flows in dynamical, 
strongly curved spacetimes  significantly extends the reach of numerical 
relativity in addressing many problems at the forefront of theoretical 
astrophysics.  We have developed and tested an evolution code for the coupled 
Einstein-Maxwell-MHD equations which combines a BSSN solver with a 
high resolution shock capturing scheme.  As 
one application, we evolve magnetized, differentially rotating neutron 
stars under the influence 
of a small seed magnetic field.  Of particular significance is the behavior 
found for hypermassive neutron stars (HMNSs), which have rest masses 
greater the mass limit allowed by uniform rotation for a given equation 
of state.  
The remnant of a binary neutron star merger is likely to be a 
HMNS.  We find that magnetic braking and the magnetorotational instability 
lead to the collapse of HMNSs and the formation of rotating black holes 
surrounded by massive, hot accretion tori and collimated magnetic field 
lines.  Such tori radiate strongly in neutrinos, and the resulting 
neutrino-antineutrino annihilation (possibly in concert with energy 
extraction by MHD effects) could provide enough energy to power 
short-hard gamma-ray bursts.  To explore the range of outcomes, we 
also evolve differentially rotating neutron stars with lower masses and 
angular momenta than the HMNS models.  Instead of collapsing, the 
non-hypermassive models form nearly uniformly rotating central objects 
which, in cases with significant angular momentum, are surrounded by 
massive tori.
\end{abstract}
%\pacs{04.25.Dm, 04.30.-w, 04.40.Dg}
\submitto{\CQG}
\maketitle

\section{Introduction}

Differential rotation is likely to be the norm in many astrophysical
settings.  For example,
differentially rotating neutron stars can form from the collapse of
massive stellar cores, which acquire rapid differential
rotation during collapse even if they are spinning uniformly at the
outset~\cite{zm97,diffrot} (see also~\cite{liu01}). Differential rotation
can also arise from the mergers of binary neutron
stars~\cite{RS99,SU00,FR00}. In these new-born, dynamically stable, 
neutron stars, magnetic
fields and/or viscosity will transport angular momentum and 
substantially change the configurations on a secular timescale. 

Some newly-formed differentially rotating neutron stars may be {\em
hypermassive}. The mass limits for non-rotating stars
[the Oppenheimer-Volkoff limit] and for rigidly rotating stars (the
{\it supramassive} limit, which is only about 20\% larger) can be
significantly exceeded by the presence of {\it differential}
rotation~\cite{BSS,MBS04}.  Mergers of binary neutron stars could lead
to the formation of such hypermassive neutron stars (HMNSs) as
remnants.  The latest binary neutron star merger
simulations in full general relativity~\cite{STUa,STUb,STUc} have
confirmed that HMNS formation is indeed a possible outcome. HMNSs
could also result from core collapse of massive stars.
Differentially rotating stars tend to approach rigid rotation when 
acted upon by processes which transport angular momentum. HMNSs,
however, cannot settle down to rigidly rotating neutron stars since 
their masses exceed the maximum allowed by rigid rotation. Thus, 
delayed collapse to a black hole and, possibly, mass loss may result. 

The merger of binary neutron stars has been proposed
for many years as an explanation of short-hard GRBs~\cite{GRB,GRB-BNS}. 
According to this 
scenario, after the merger, a stellar-mass black hole is formed,
surrounded by hot accretion torus containing $\sim 1$--$10\%$ of 
the total mass of the system.  The GRB fireball is then powered 
by energy extracted from this system, either by MHD processes or
emission from neutrino-antineutrino annihilation.  The viability of 
this model depends on the presence of a significantly massive accretion 
disk after the collapse of the remnant 
core, which in turn depends on the mechanism driving the collapse.

Though magnetic fields likely play a significant role in the evolution 
of HMNSs, the numerical tools needed to study this problem have not been
available until recently.  In particular, the evolution of magnetized HMNSs
can only be determined by solving the coupled Einstein-Maxwell-MHD equations
self-consistently in full general relativity.  
Recently, Duez {\it et al.}~\cite{DLSS2} and Shibata
and Sekiguchi~\cite{SS} independently developed codes designed 
to do such calculations 
for the first time (see also~\cite{valencia:fn,byu}).  
The first simulations of magnetized 
hypermassive neutron star collapse (assuming both axial and equatorial 
symmetry) were reported in~\cite{DLSSS1}, and the implications of these results 
for short GRBs were presented in~\cite{GRB2}.  These simulations proved that 
the amplification of small seed magnetic fields by a combination of magnetic 
winding and the magnetorotational instability (MRI)~\cite{MRI0,MRIrev} is sufficient to trigger 
collapse in hypermassive stars on the Alfv\'en timescale, %$t_A \sim R/v_A$, 
confirming earlier predictions~\cite{BSS,Shapiro}.   

We also compare the results 
for hypermassive stars with the evolution of two differentially rotating 
models below the supramassive limit in order to highlight the qualitatively 
different physical effects which arise in the evolution. Given a fixed equation 
of state (EOS), the sequence of 
uniformly rotating stars with a given rest mass has a maximum angular momentum 
$J_{\rm max}$. A nonhypermassive star having angular momentum $J>J_{\rm max}$ 
is referred to as an ``ultraspinning'' star. We perform simulations 
on the MHD evolution of two nonhypermassive stars -- one is ultraspinning 
and the other is not; we refer to the latter as ``normal.''  
Instead of collapsing, they evolve to a 
new equilibrium state after several Alfv\'en times. The normal
star settles down to a uniformly rotating configuration. In contrast, the 
ultraspinning star settles down to 
a nearly uniformly rotating central core, surrounded by 
a differentially rotating torus. 

The key subtlety in all of these simulations is that the wavelengths 
of the MRI modes 
must be well resolved on the computational grid.  Since this wavelength is 
proportional to the magnetic field strength, it becomes very difficult to 
resolve for small seed fields.  However, the simulations reported here succeed 
in resolving the MRI. 
In what follows, we assume geometrized units such that $G=c=1$.

\section{Initial Models}
\label{models}
We consider three representative differentially rotating 
stars, which we call ``A'', ``B1'', and ``B2.'' 
Their properties are listed in Table~\ref{startable}. 
Star~A is hypermassive while stars~B1 and B2 are not. 
These configurations are all dynamically stable.
All three models are constructed using a $\Gamma=2$ polytropic EOS, 
$P=K\rho_0^{\Gamma}$, where $P$, $K$, and $\rho_0$ are the pressure, 
polytropic constant, and rest-mass density, respectively.  
The rest mass of star~A 
exceeds the supramassive limit by 46\%, while the rest masses of stars~B1 
and B2 are below the supramassive limit. The angular momentum of star~B1 
exceeds the maximum angular momentum ($J_{\rm max}$) for a rigidly 
rotating star with the same rest mass and EOS, whereas star~B2 has 
angular momentum $J<J_{\rm max}$. Thus, star~B1 is ``ultraspinning,''
while star~B2 is ``normal.''  These models may be scaled to any desired 
physical mass by adjusting the value of $K$~\cite{CST}.

\begin{table*}
\caption{Initial Models.}
\begin{center}
\begin{tabular}{c c c c c c c c}
\hline 
\hline
\vspace{0.03in}
Case &  \small{$M_0/M_{0,\rm{TOV}}$${}^{\rm a}$} & 
\small{$M_0/M_{0,\rm{sup}}$${}^{\rm b}$}  & 
 $R_{\rm eq}/M$${}^{\rm c}$ & $J/M^2$${\ }^{\rm d}$ 
& $T_{\rm rot}/|W|$${}^{\rm e}$ & $\Omega_{\rm eq}/\Omega_c$${}^{\rm f}$ \\
\hline\hline
  A  & 1.69 & 1.46 & 4.48  &  1.0  & 0.249 & 0.33  \\
\hline
  B1 & 0.99 & 0.86 & 8.12  &  1.0  & 0.181 & 0.40  \\
\hline
 B2 & 0.98 & 0.85 & 4.84  &  0.38  & 0.040 & 0.34  \\
\hline \hline
\end{tabular}
\end{center}
\vskip 12pt
\begin{minipage}{13cm}
\raggedright
${}^{\rm a}$ {The ratio of the rest mass $M_0$ to the TOV rest mass limit for the given EOS.}
\\
${}^{\rm b}$ {The ratio of the rest mass $M_0$ to the rest mass limit for uniformly rotating
stars of the given EOS (the supramassive limit).  If this ratio is greater than unity, 
the star is hypermassive.}
\\
${}^{\rm c}$ {The equatorial coordinate radius $R_{\rm eq}$ normalized by 
the ADM mass.}
\\
${}^{\rm d}$ {The ratio of the angular momentum $J$ to $M^2$ (the angular momentum 
parameter).}
\\
${}^{\rm e}$ {The ratio of the rotational kinetic energy to the gravitational 
binding energy.}
\\
${}^{\rm f}$ {The ratio of the angular velocity at the equator to the 
central angular velocity.}
\end{minipage}
\label{startable}
\end{table*}

Following previous papers (e.g, \cite{CST,BSS,SBS,DLSS}), we choose the 
initial rotation law $u^0 u_{\varphi}=A^2(\Omega_c-\Omega)$, where $u^{\mu}$ is 
the four-velocity, $\Omega_c$ is the angular velocity along the rotational axis,
and $\Omega \equiv u^{\varphi}/u^0$ is the angular velocity. 
The constant $A$ has units of length and determines the steepness of the
differential rotation. For these models, $A$ is set equal to the
coordinate equatorial radius $R_{\rm eq}$. The corresponding values of 
$\Omega_{\rm eq}/\Omega_c$ are shown in Table~\ref{startable} 
(where $\Omega_{\rm eq}$ is the angular velocity at the equatorial 
surface).  
%Stars~A and B1 
%rotate very rapidly and are highly flattened due to centrifugal force, while
%star~B2 rotates comparatively slowly.

We must also specify initial conditions for the magnetic field.  We choose
to add a weak poloidal magnetic field to the equilibrium model 
by introducing a vector potential of the following form
$A_{\varphi}= \varpi^2 {\rm max}[A_b (P-P_{\rm cut}), 0]$, 
where the cutoff $P_{\rm cut}$ is 4\% of the maximum pressure, and $A_b$ 
is a constant which determines the initial strength of
the magnetic field. We characterize the strength of the initial magnetic 
field by $C\equiv {\rm max}(B_{(u)}^2/(8\pi P))$, i.e.\ the maximum 
value on the grid 
of the ratio of the magnetic energy density to the pressure (where 
$B_{(u)}^{\mu}$ refers to the magnetic field in the comoving frame 
of the fluid).  We choose
$A_b$ such that $C\sim 10^{-3}$--$10^{-2}$. We have verified 
that such small initial magnetic fields introduce negligible violations
of the Hamiltonian and momentum constraints in the initial data.

\section{Numerical Methods}
\label{sec:methods}
Duez {\it et al.}~\cite{DLSS2} and Shibata and Sekiguchi~\cite{SS} have
independently developed new codes to evolve magnetized fluids in 
dynamical spacetimes
by solving the Einstein-Maxwell-MHD system of equations self-consistently.
Several tests have been performed
with these codes, including MHD shocks, nonlinear MHD wave propagation, 
magnetized Bondi
accretion, MHD waves induced by linear gravitational waves, and magnetized
accretion onto a neutron star.  Details of our techniques for evolving the
Einstein-Maxwell-MHD system as well as tests can be found in~\cite{DLSS2,SS}. 
We have performed several simulations for identical 
initial data using both codes and 
found that the results are essentially the same.
The simulations presented here assume axial and equatorial 
symmetry. We adopt 
the Cartoon method~\cite{cartoon} for evolving the BSSN equations~\cite{BSSN}, and use 
cylindrical coordinates for evolving the induction and MHD equations. In this 
scheme, the coordinate $x$ is identified with the cylindrical 
radius $\varpi$, and the $y$-direction corresponds to the azimuthal 
direction. 
%As in many hydrodynamic simulations, we add a tenuous, uniform-density 
%``atmosphere'' to cover the computational grid outside the star. 

\section{Results}
\label{sec:results}
\subsection{Star~A}
\label{starA}

We have performed simulations on star~A with initial field 
strength given by $C=2.5 \times 10^{-3}$. We use a uniform grid with size 
$(N,N)$ in cylindrical coordinates $(\varpi,z)$, which covers the region 
$[0,L]$ in each direction, where $L=4.5R_{\rm eq}$. For star~A, 
$R_{\rm eq}=4.5M=18.6~{\rm km} (M/2.8M_{\odot})$. To check the 
convergence of our numerical results, 
we perform simulations with four different grid resolutions: 
$N=$ 250, 300, 400 and 500. Unless otherwise stated, 
all results presented in the following subsections are from  
the simulation data with resolution $N=500$.  

\subsubsection{General features of the evolution}

\begin{figure*}
\begin{center}
\epsfxsize=1.8in
\leavevmode
\hspace{-0.7cm}\epsffile{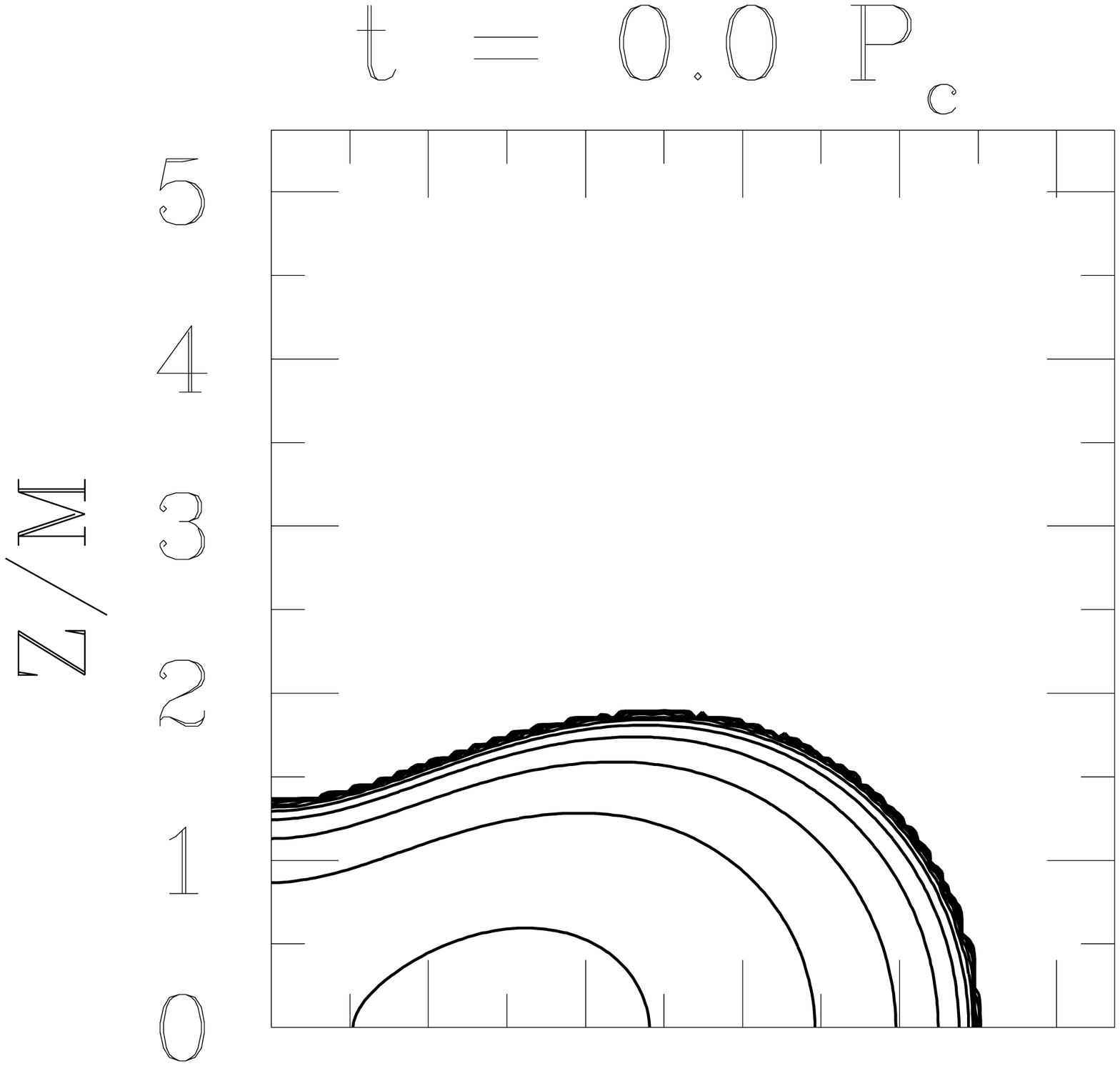}
\epsfxsize=1.8in
\leavevmode
\hspace{-0.5cm}\epsffile{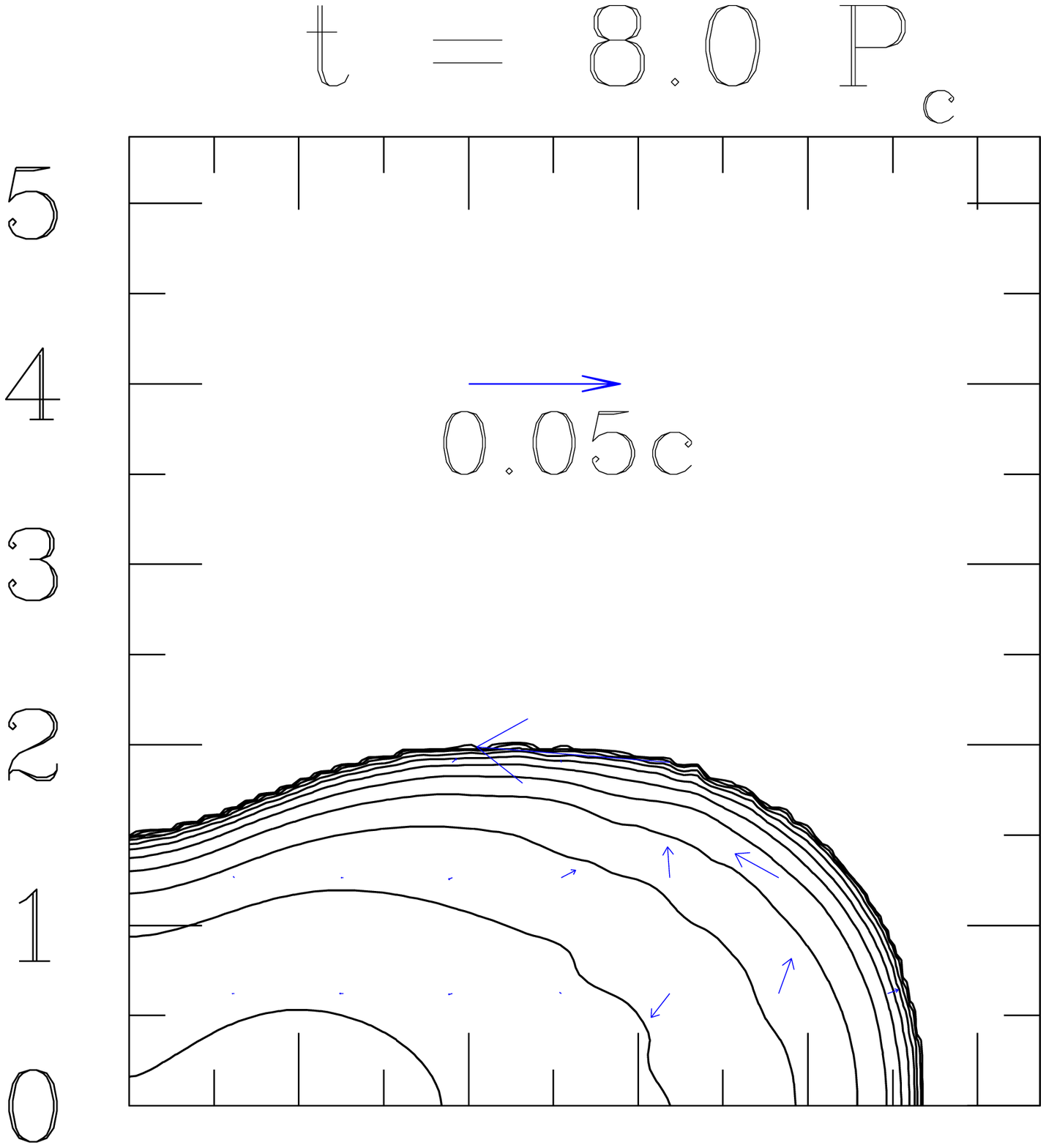}
\epsfxsize=1.8in
\leavevmode
\hspace{-0.5cm}\epsffile{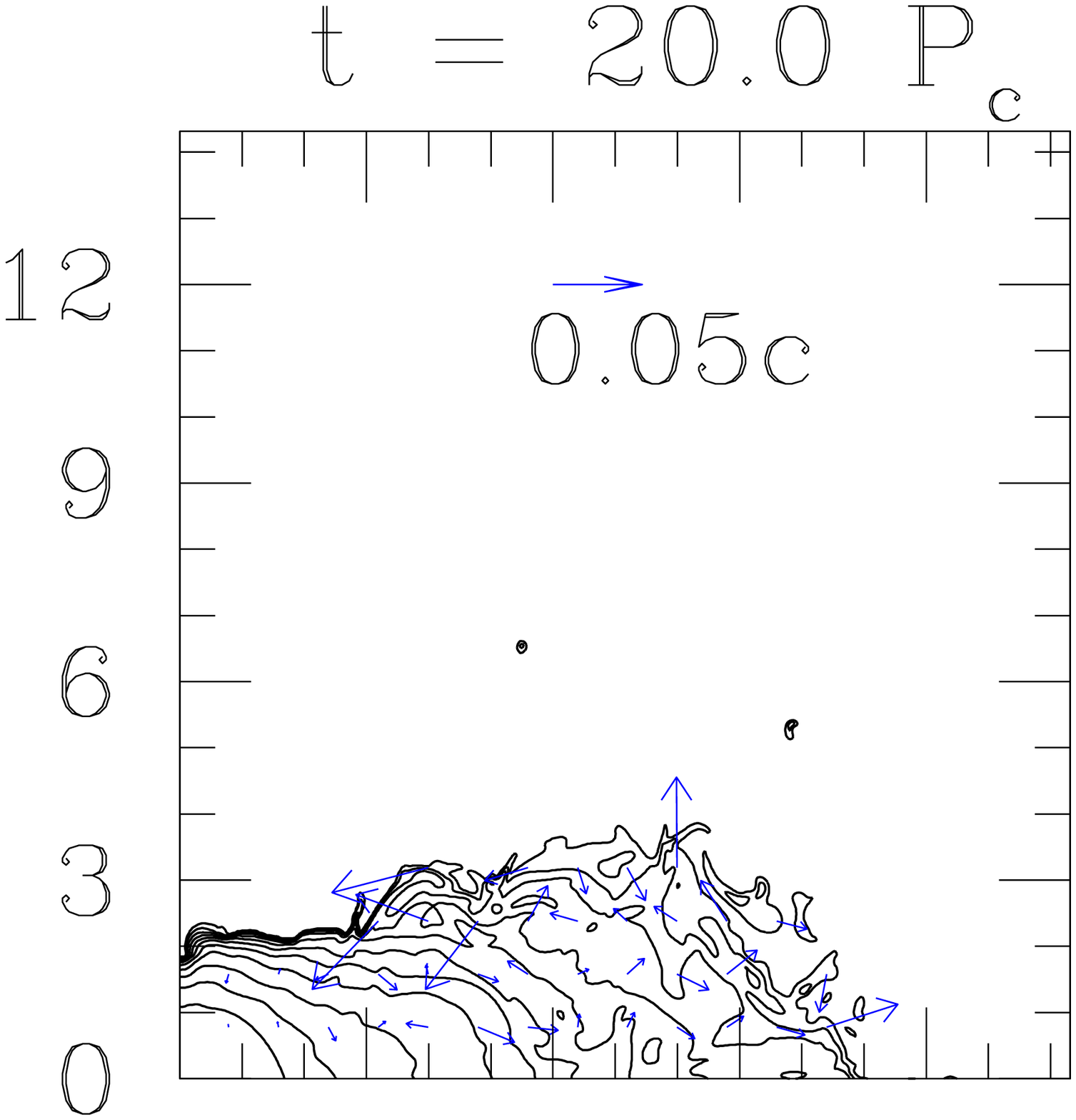} \\
\vspace{-0.5cm}
\epsfxsize=1.8in
\leavevmode
\hspace{-0.7cm}\epsffile{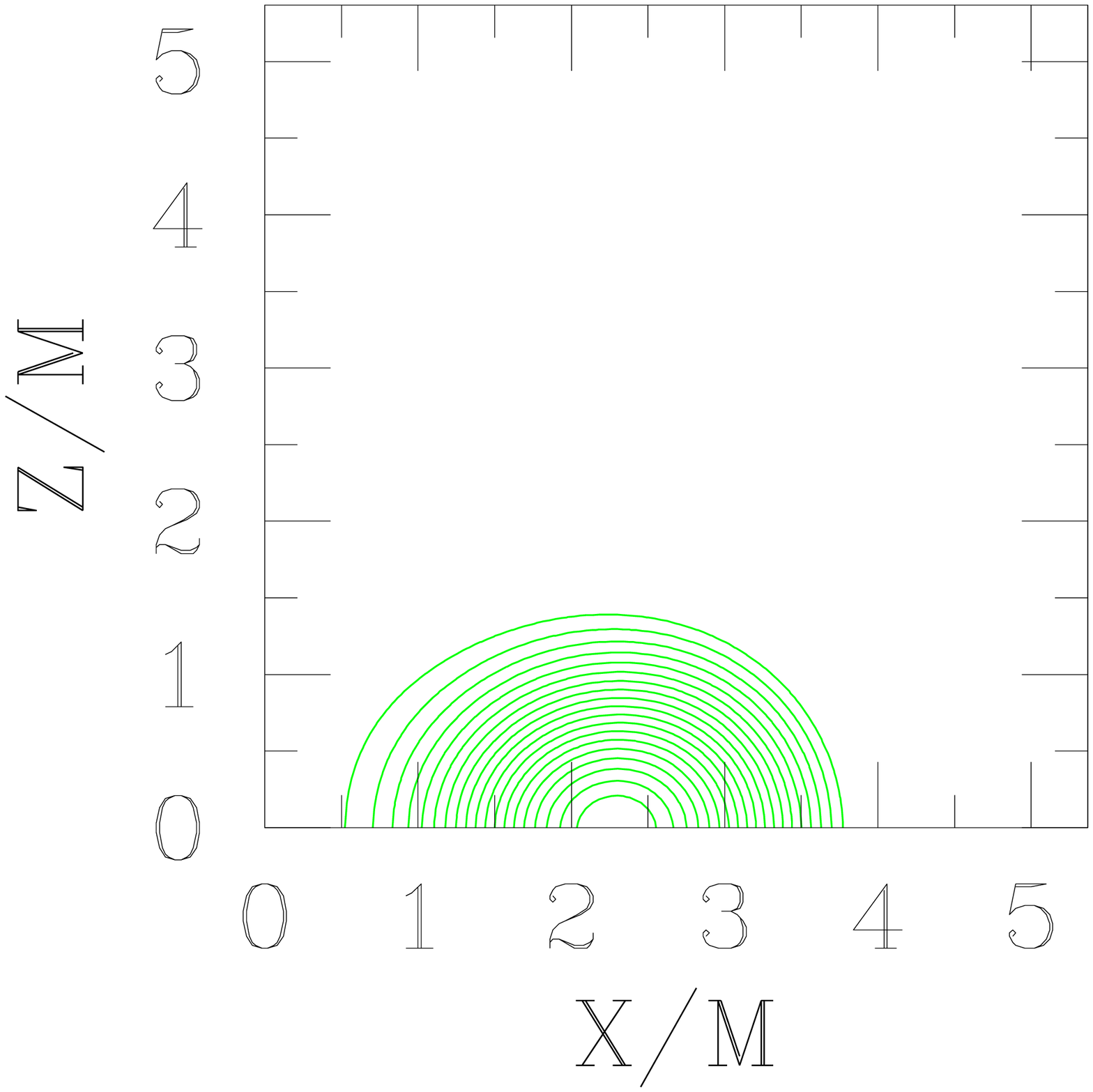}
\epsfxsize=1.8in
\leavevmode
\hspace{-0.5cm}\epsffile{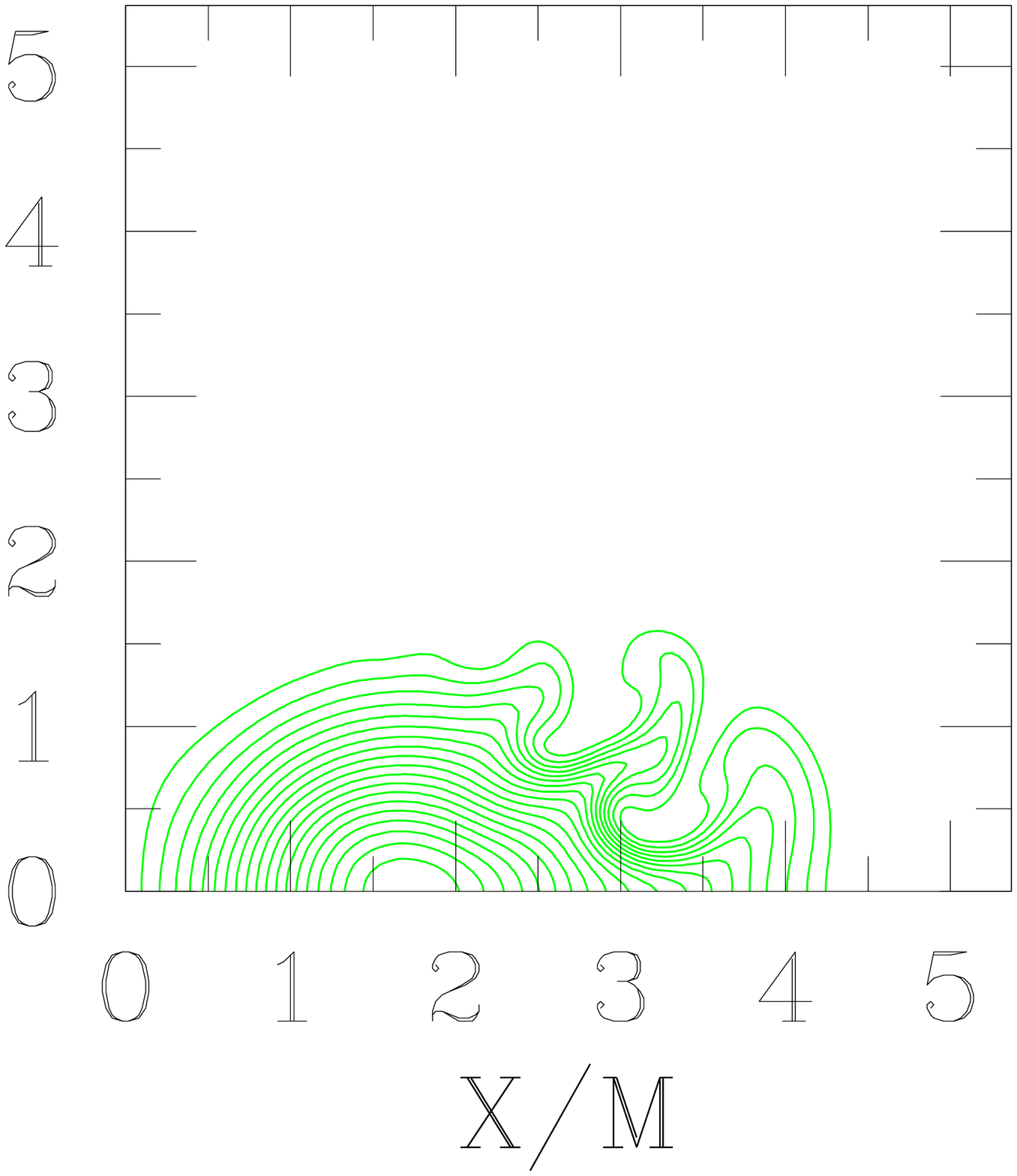}
\epsfxsize=1.8in
\leavevmode
\hspace{-0.5cm}\epsffile{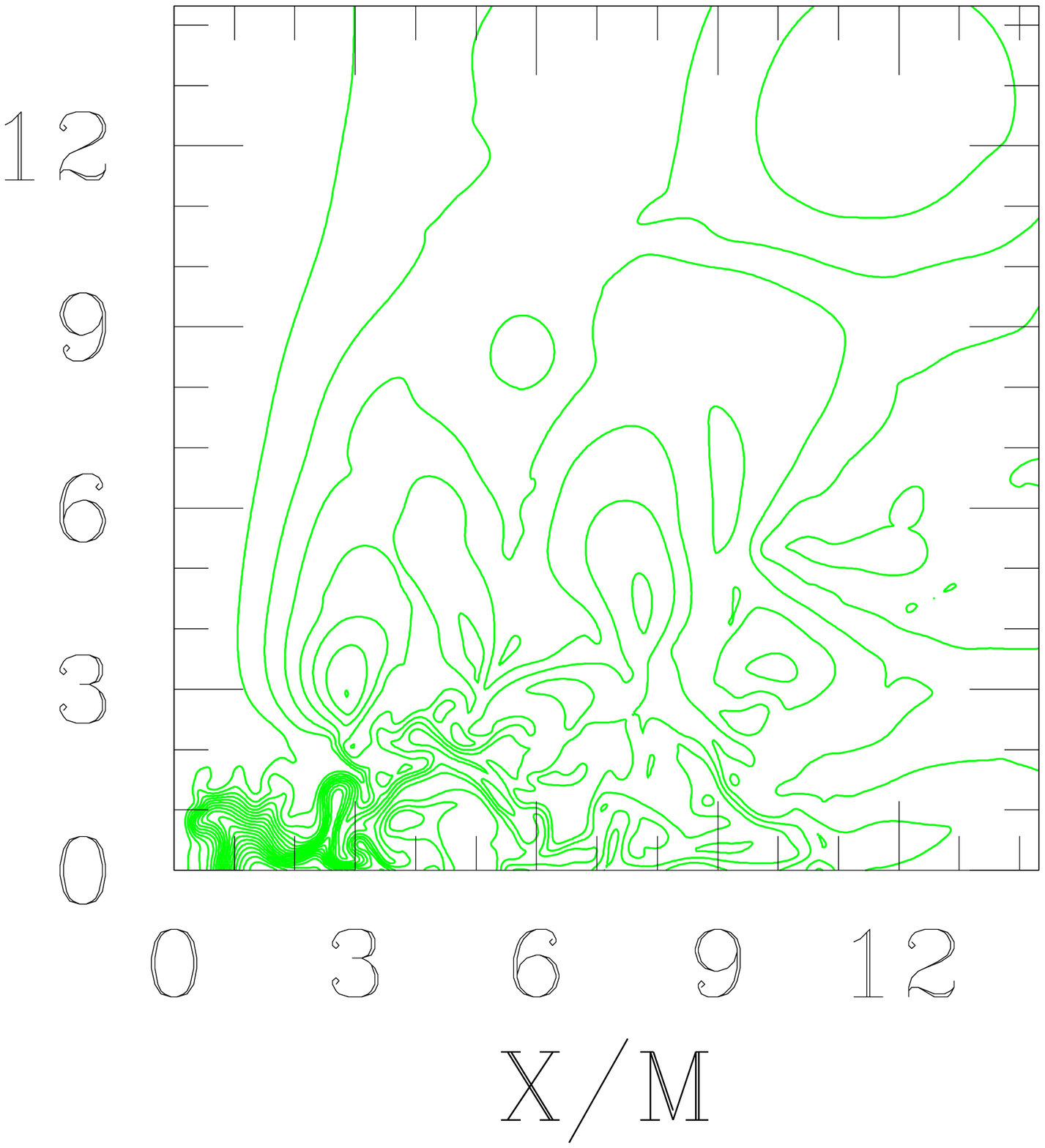}
\end{center}

\begin{center}
\epsfxsize=1.8in
\leavevmode
\hspace{-0.7cm}\epsffile{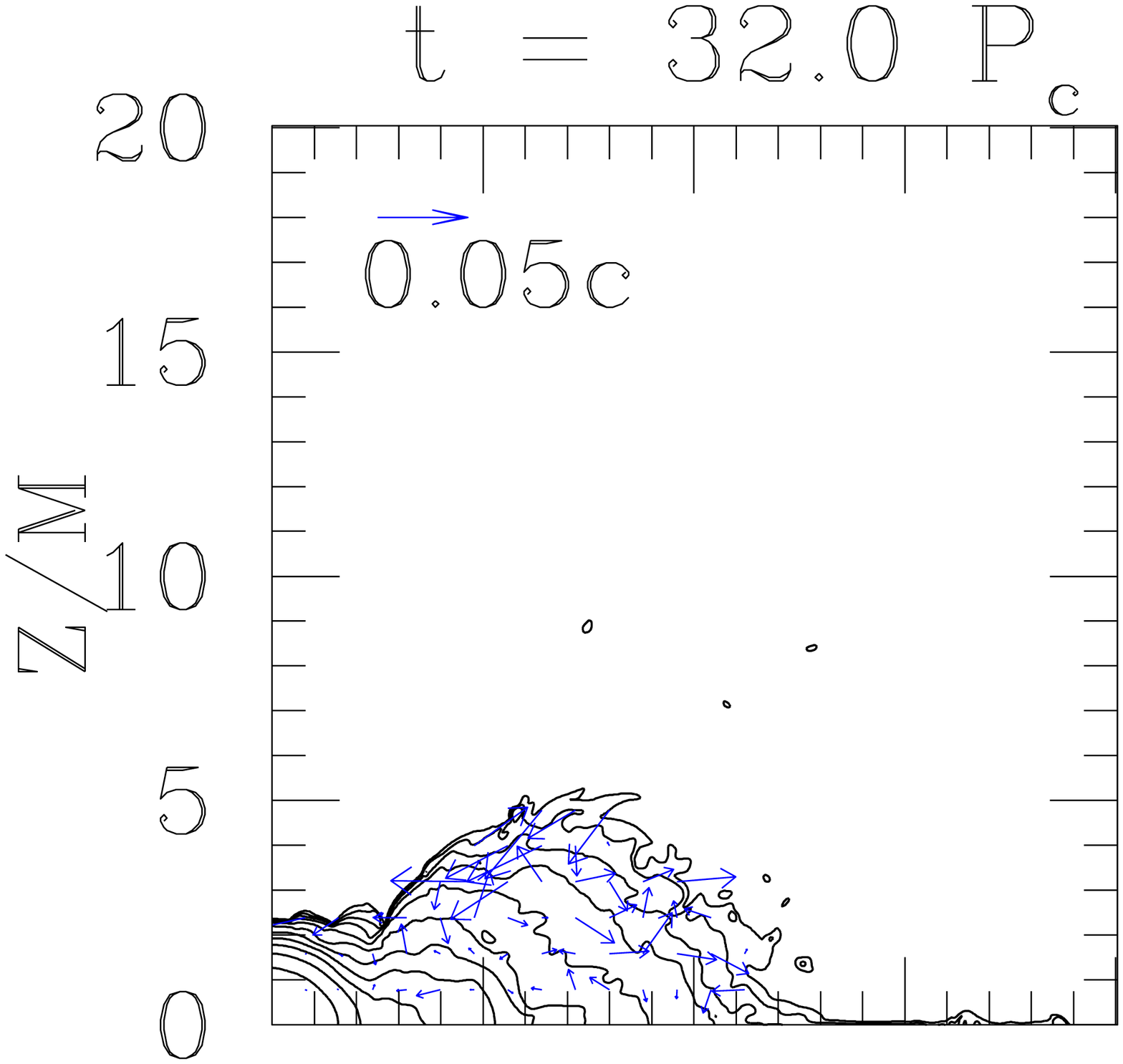}
\epsfxsize=1.8in
\leavevmode
\hspace{-0.5cm}\epsffile{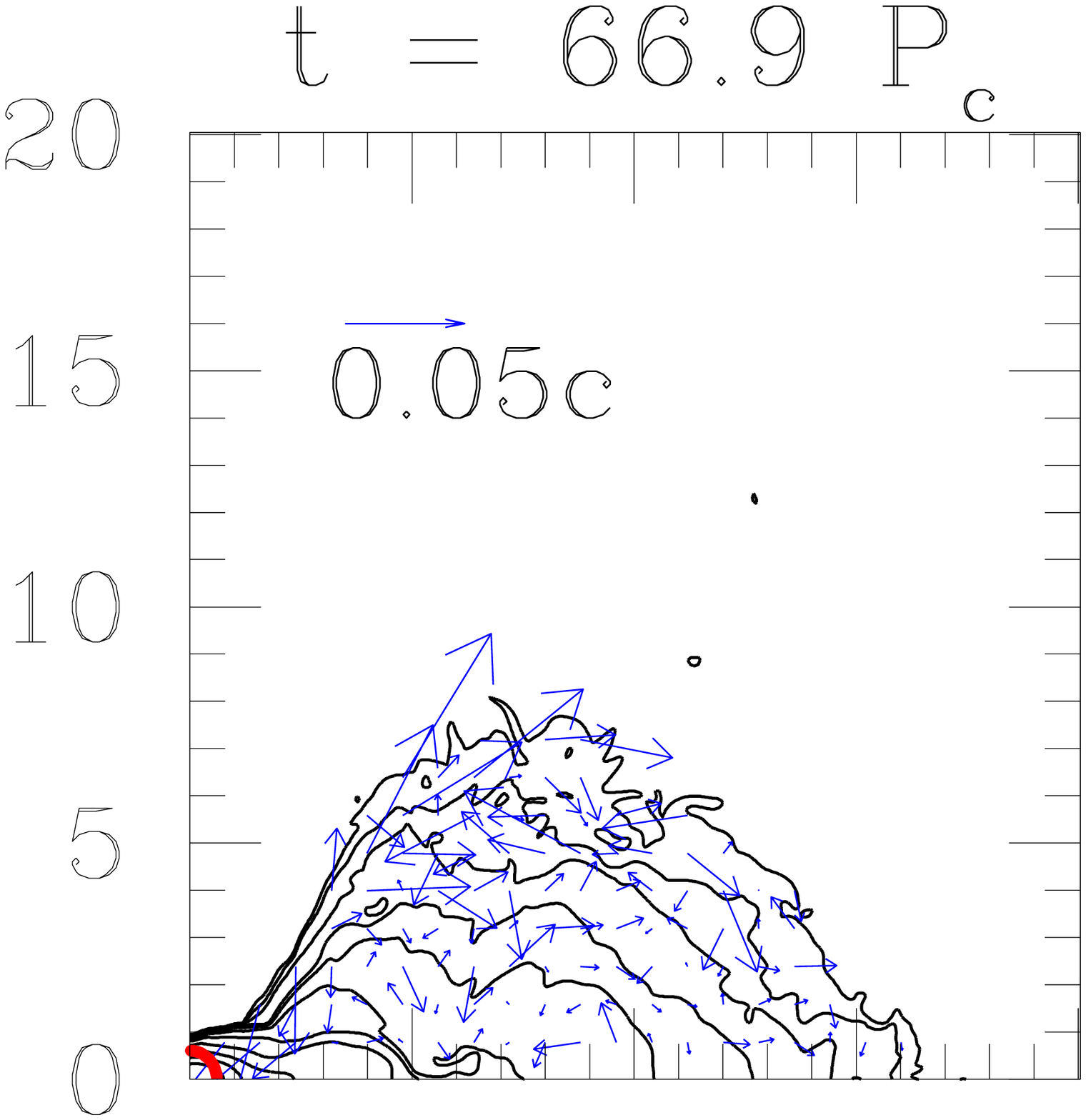}
\epsfxsize=1.8in
\leavevmode
\hspace{-0.5cm}\epsffile{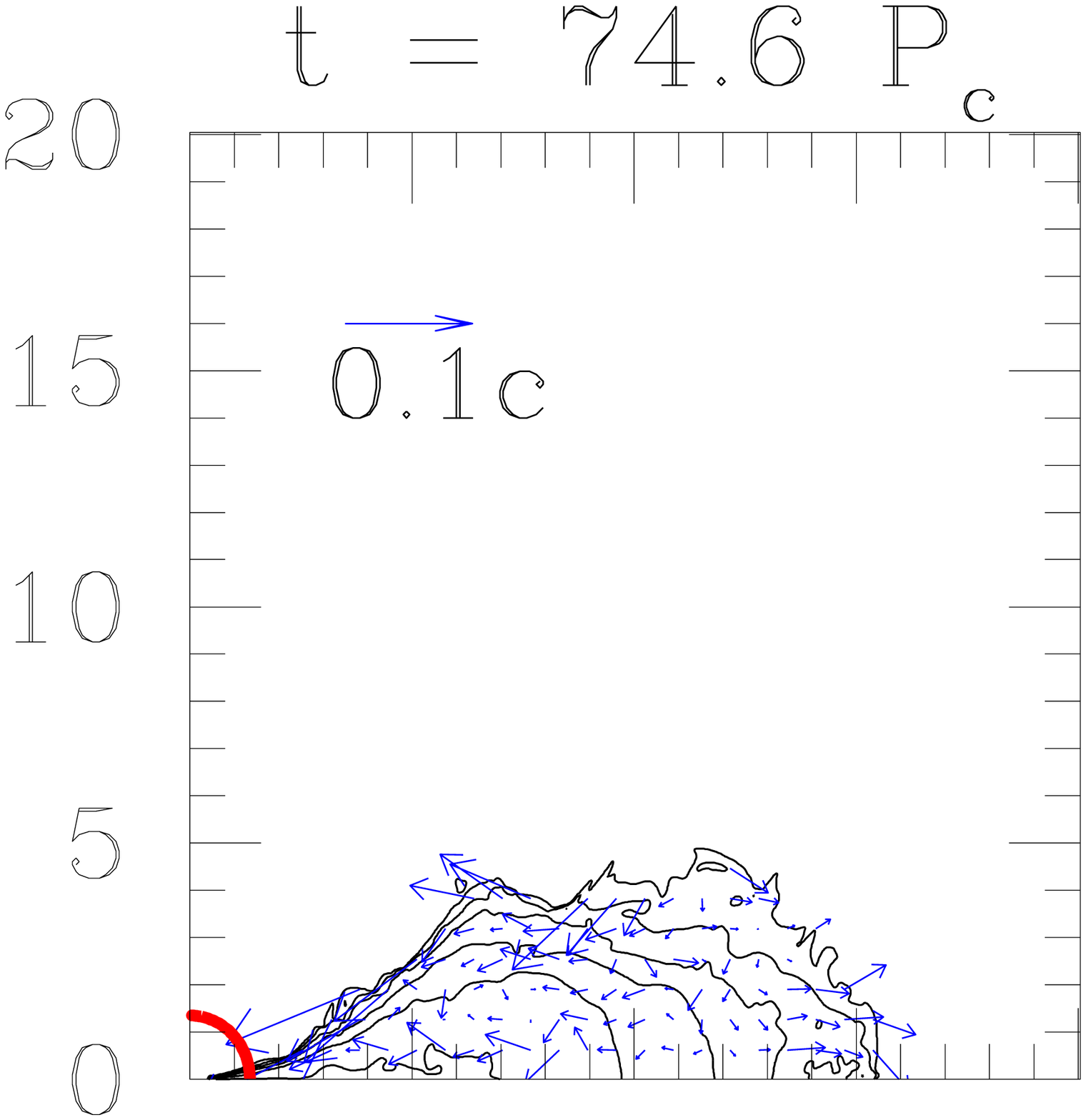} \\
\vspace{-0.5cm}
\epsfxsize=1.8in
\leavevmode
\hspace{-0.7cm}\epsffile{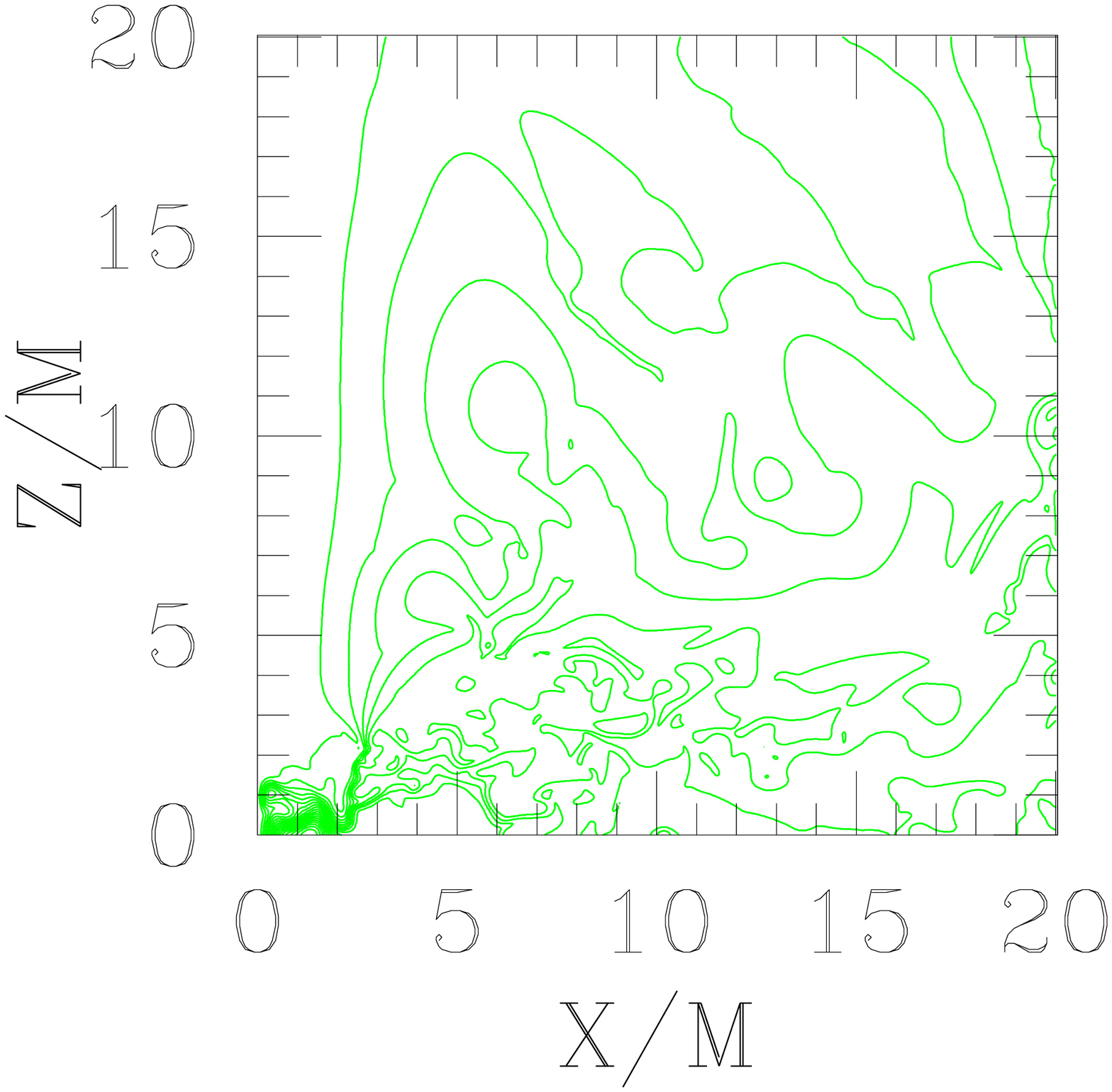}
\epsfxsize=1.8in
\leavevmode
\hspace{-0.5cm}\epsffile{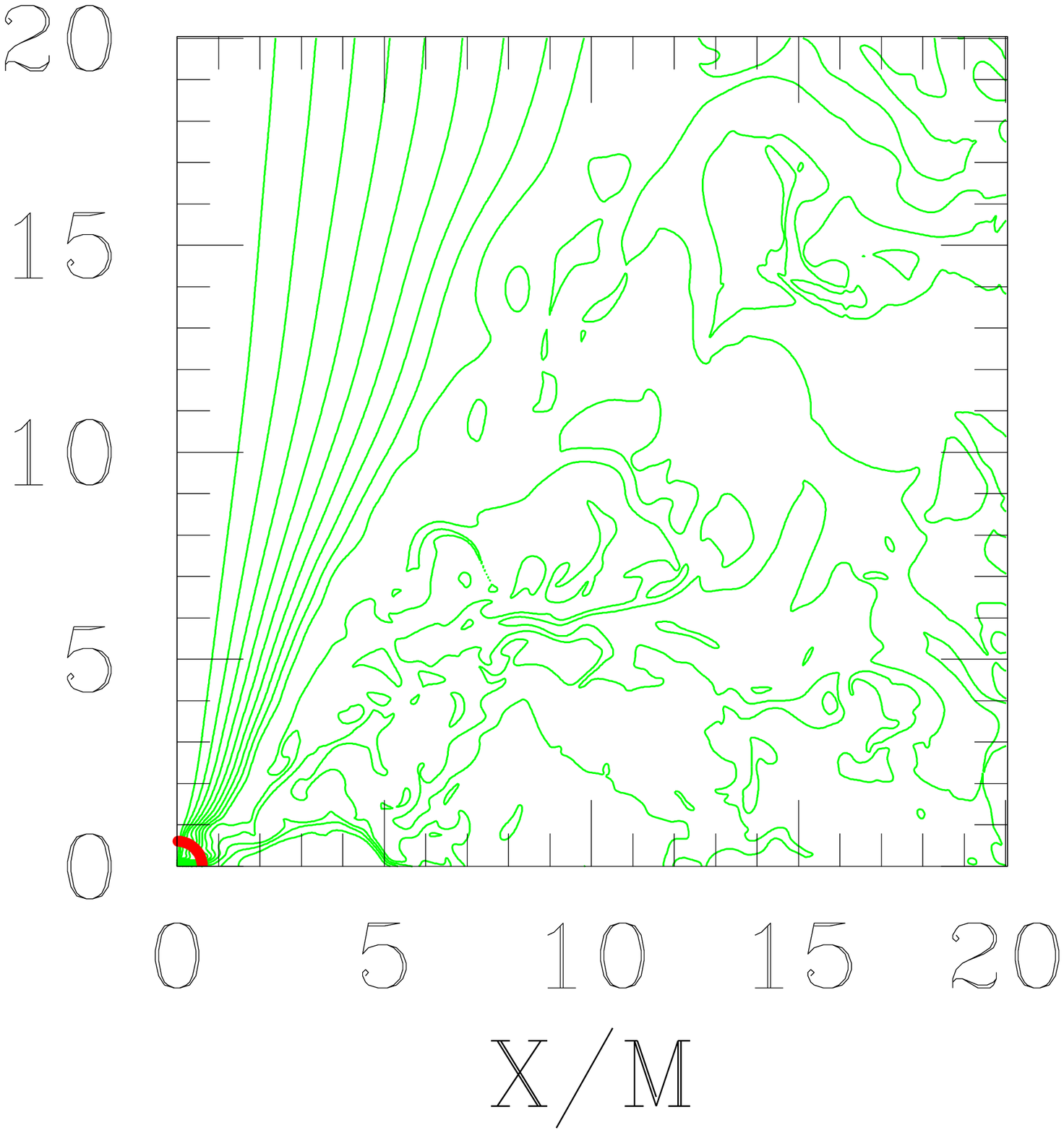}
\epsfxsize=1.8in
\leavevmode
\hspace{-0.5cm}\epsffile{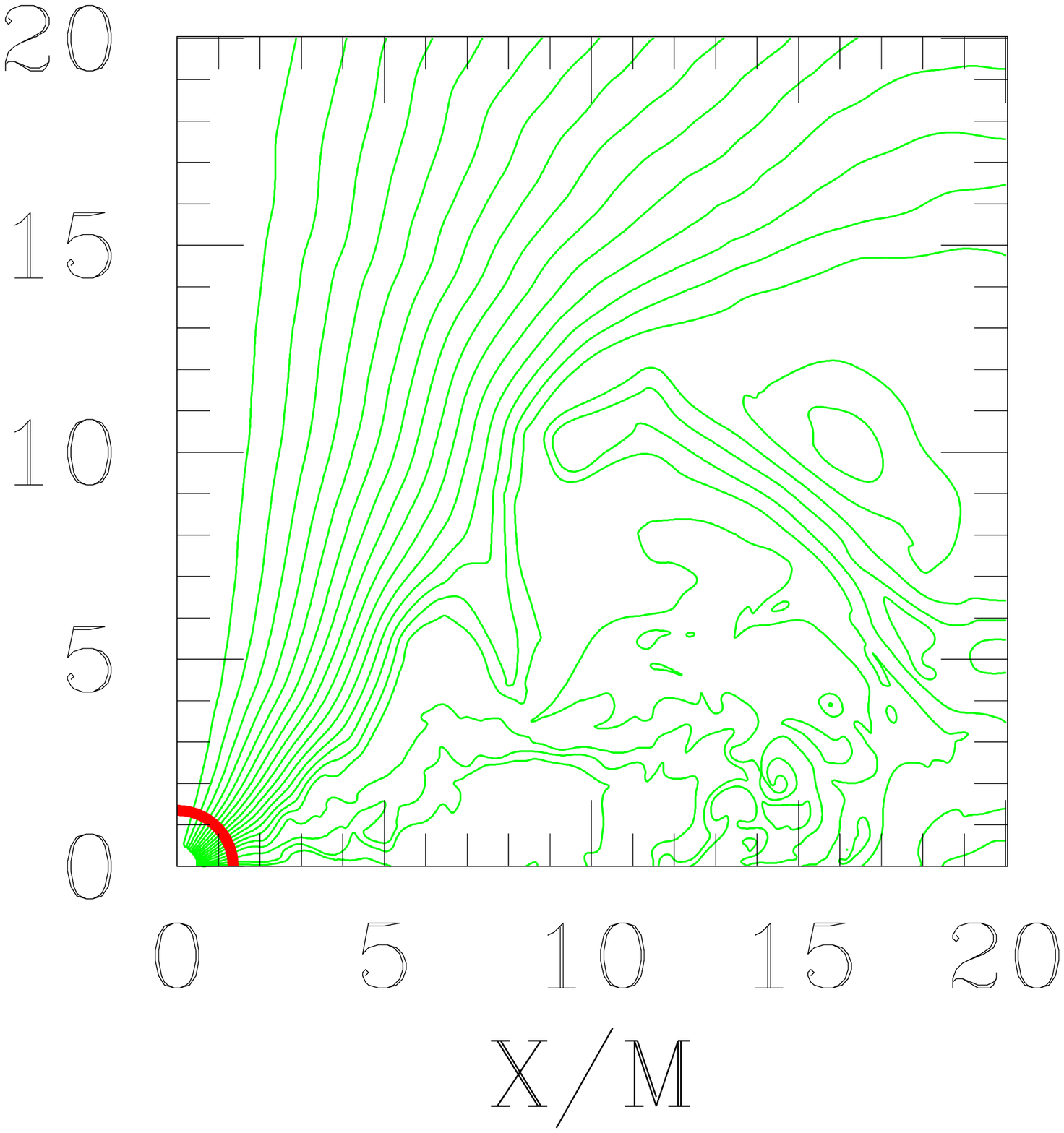}
\caption{Snapshots of  rest-mass density contours and poloidal 
magnetic field lines for star~A at selected times. 
The first and third rows show snapshots of the rest-mass density
contours and velocity vectors on the meridional plane. The second and fourth
rows show the corresponding field lines (lines of constant $A_{\varphi}$)
for the poloidal magnetic field at the same times.
The density contours are drawn for $\rho_0/\rho_{\rm max}(0)=
10^{-0.36 i - 0.09}~(i=0$--10), where $\rho_{\rm max}(0)$ is the 
maximum rest-mass density at $t=0$. 
The field lines are drawn for $A_{\varphi} = A_{\varphi,\rm min}
+ (A_{\varphi,\rm max} - A_{\varphi,\rm min}) i/20~(i=1$--19),
where $A_{\varphi,\rm max}$ and $A_{\varphi,\rm min}$ are the maximum
and minimum values of $A_{\varphi}$, respectively, at the given time.
The thick solid curves denote the apparent horizon. 
In the last panel, the field lines are terminated inside the black hole
at the excision boundary.
\label{fig:StarA_contours}}
\end{center}
\end{figure*}

Figure~\ref{fig:StarA_contours} shows snapshots of the density contours 
and poloidal magnetic field lines (lines of constant $A_{\varphi}$) in 
the meridional plane. 
In the early phase of the evolution, the frozen-in poloidal magnetic fields 
lines are wound up by the differentially rotating matter, creating 
a toroidal field which grows linearly in 
time (see Fig.~\ref{rescomp}d). 
When the magnetic field becomes sufficiently strong, magnetic stresses 
act back on 
the fluid, causing a redistribution of angular momentum. The core 
of the star contracts while the outer layers expand. As shown in 
Fig.~\ref{rescomp}c, the effect of the MRI becomes evident for $t \agt 
6P_c$.  This is manifested as a sudden increase in 
the maximum value of $|B^x|$ ($\equiv |B^{\varpi}|$), which  
grows exponentially for a short period (about one $e$-folding) before 
saturating. 
The effect of the MRI is also visible in Fig.~\ref{fig:StarA_contours}, 
where we see distortions in the poloidal field lines. 
The amplitude of 
the toroidal field begins to decrease after $t \agt 20P_c \sim t_A$ 
(see Fig.~\ref{rescomp}d) and the core 
of the star becomes less differentially rotating.

The combined effects of magnetic braking and the MRI eventually trigger 
gravitational collapse to a black hole at $t \approx 66P_c \approx 
36 (M/2.8M_{\odot})~{\rm ms}$, where $P_c$ is the initial central
rotation period. A 
collimated magnetic field forms near the polar 
region at this time (see Fig.~\ref{fig:StarA_contours}). However, 
a substantial amount of toroidal field is still present. 
Without black hole excision, the 
simulation becomes inaccurate soon after the formation of the apparent 
horizon because of grid stretching. To follow the subsequent 
evolution, a simple excision technique is 
employed~\cite{alcubierre,excision}. 
We are able to track the evolution for another $300M \approx 8 P_c$. 
We find that not all of the matter promptly falls into the black hole. 
The system settles down to a quasiequilibrium state consisting of a 
black hole surrounded by a hot torus and a collimated magnetic 
field near the polar region (see the panels corresponding to time 
$t=74.6P_c$ in Fig.~\ref{fig:StarA_contours}). 

\subsubsection{Resolution study}

\begin{figure}
\epsfxsize=2.5in
\epsffile{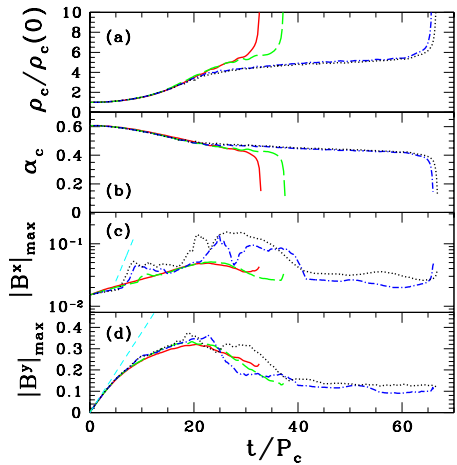}
\caption{Evolution of the central rest-mass density $\rho_c$, central lapse
$\alpha_c$, and
maximum values of $|B^x|$ and $|B^y|$.
$|B^x|_{\rm max}$ and $|B^y|_{\rm max}$ are plotted in units of 
$\sqrt{\rho_{\max}(0)}$. 
The solid, long-dashed, dot-dash, 
and dotted curves denote
the results with $N$=250, 300, 400, and 500 respectively.
The dashed  
line in (c) represents an approximate slope $\omega=0.18/P_c$ for the 
exponential growth rate of the MRI, $\delta B^x \propto e^{\omega t}$. 
The dashed line in (d) represents the predicted
growth of $|B^y|_{\rm max}$ at early times from linear theory.
\label{rescomp}}
%\end{center}
\end{figure}

Four simulations were performed with different resolutions 
(see Fig.~\ref{rescomp}): $N=$ 250, 300, 400 and 500. 
We find that the results converge approximately when $N\agt 400$. 
On the other hand, results are far from convergent for $N \alt 300$. 
Most importantly, the effect of the MRI is not captured in 
the two lowest resolution runs, as shown by the 
behavior of $|B^x|_{\rm max}$ in Fig.~\ref{rescomp}. This is 
because the wavelength of the fastest growing MRI mode ($\lambda_{\rm max}$) 
is not well-resolved for low resolutions. We find that we need a resolution 
$\Delta/\lambda_{\rm max} \alt 0.14$ ($N \agt 400$) in order to 
resolve the MRI modes.  
In contrast, the growth of $B^y$ by magnetic winding is resolved
for all four resolutions.  The straight dashed line in Fig.~\ref{rescomp}d 
corresponds to the prediction of linear theory for the growth rate due 
to winding.  
This slope agrees with the actual growth of $|B^y|_{\rm max}$ in the early 
(magnetic winding) phase of the simulation, but as back-reaction 
(magnetic braking) becomes important, the toroidal field begins to saturate.
(See~\cite{bigpaper} for a discussion of the MRI and winding effects
in linear theory.)

\subsubsection{Evolution with excision}
\label{sec:SA-excision}

\begin{figure}
\vspace{-4mm}
\begin{center}
\epsfxsize=2.5in
\leavevmode
\epsffile{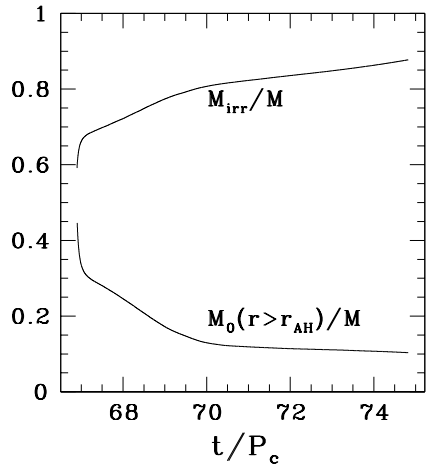}
\caption{Evolution of the irreducible mass 
and the total rest-mass outside the apparent horizon.  (Here,
$r_{\rm AH}$ is the local coordinate radius of the apparent horizon.)
\label{excision}}
\end{center}
\end{figure}

Soon after the formation of the apparent horizon, the simulation
becomes inaccurate due to grid stretching and an excision technique is 
required to follow the subsequent evolution.  For details on
our excision techniques, see~\cite{alcubierre,ybs02,excision}.
During the excision
evolution, we track the irreducible mass of the black hole by computing
the area of the apparent horizon $\mathcal{A}_{\rm AH}$ and using  
$M_{\rm irr} \approx \sqrt{\mathcal{A}_{\rm AH}/16\pi}$.  The irreducible 
mass and the total rest mass outside the apparent horizon are 
shown in Fig.~\ref{excision}.  The total ADM mass of the final state 
system, consisting of a BH surrounded by a massive accretion torus, is 
well defined.  In contrast, there is no rigorous definition for the mass 
of the black hole itself.  We obtain a rough estimate following the
steps described in~\cite{bigpaper}.  We find that
$M_{\rm hole } \sim 0.9 M$, where $M$ is the total ADM mass of the 
system, and $J_{\rm hole}/M_{\rm hole}^2 \sim 0.8$.

The black hole grows at an initially rapid rate following its formation. 
However, the accretion rate $\dot M_0$ gradually 
decreases and the black hole settles down to a quasi-equilibrium state. 
By the end of the simulation, $\dot M_0$ has decreased
to a steady rate of $\approx 0.01 M_0/P_c$, giving an accretion timescale of 
$\sim 10$--$20P_c \approx 5$--$10~{\rm ms} (M/2.8M_{\odot})$.  
Also, we find that the specific internal thermal energy in the torus 
near the surface is substantial because of shock heating.  The possibility
that this sort of system could give rise to a GRB is discussed 
in~\cite{GRB2}.

\begin{figure*}
\begin{center}
\epsfxsize=2.4in
\leavevmode
\hspace{-0.7cm}\epsffile{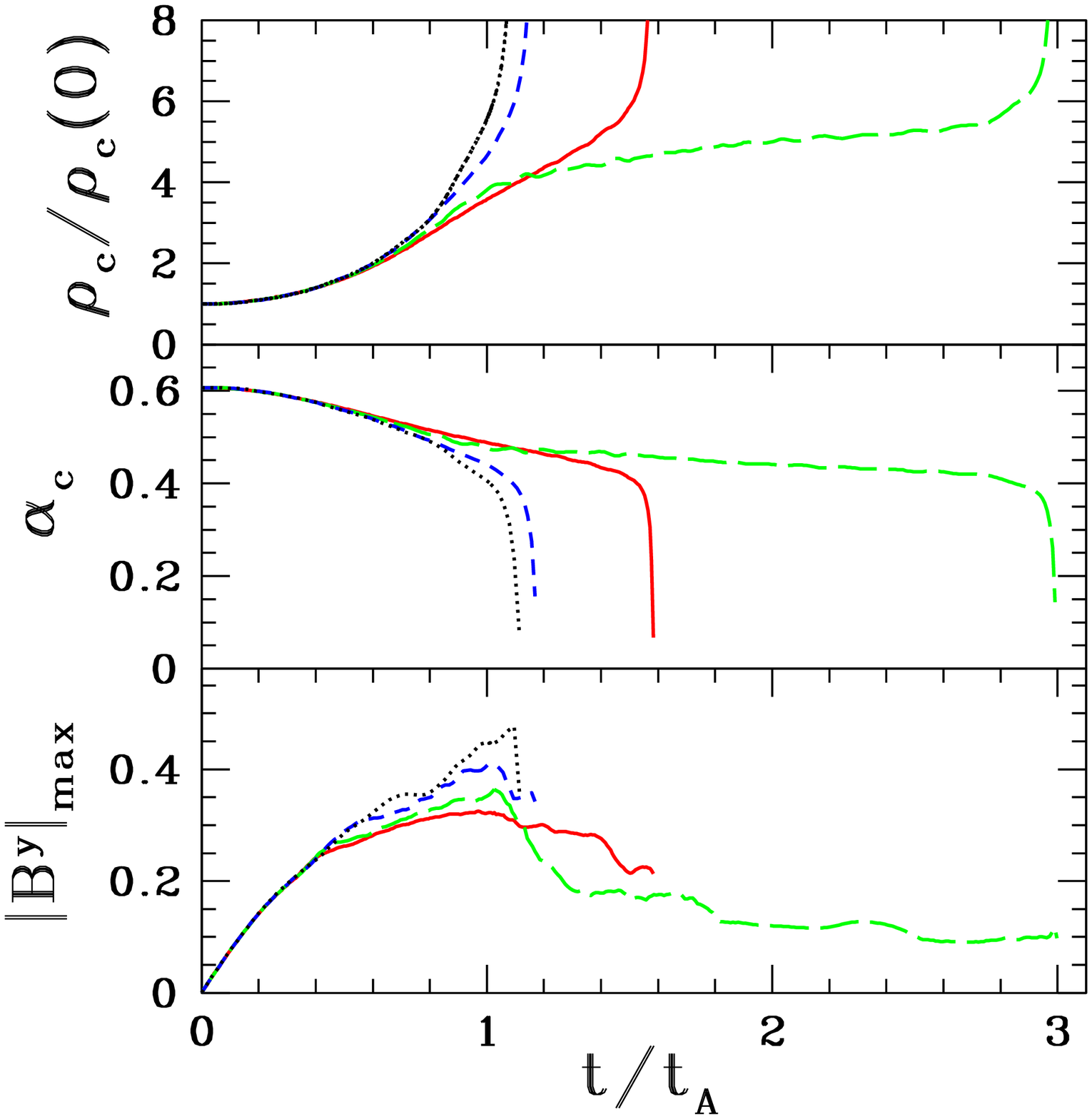}
\epsfxsize=2.5in
\leavevmode
\vspace{-0.25cm}\epsffile{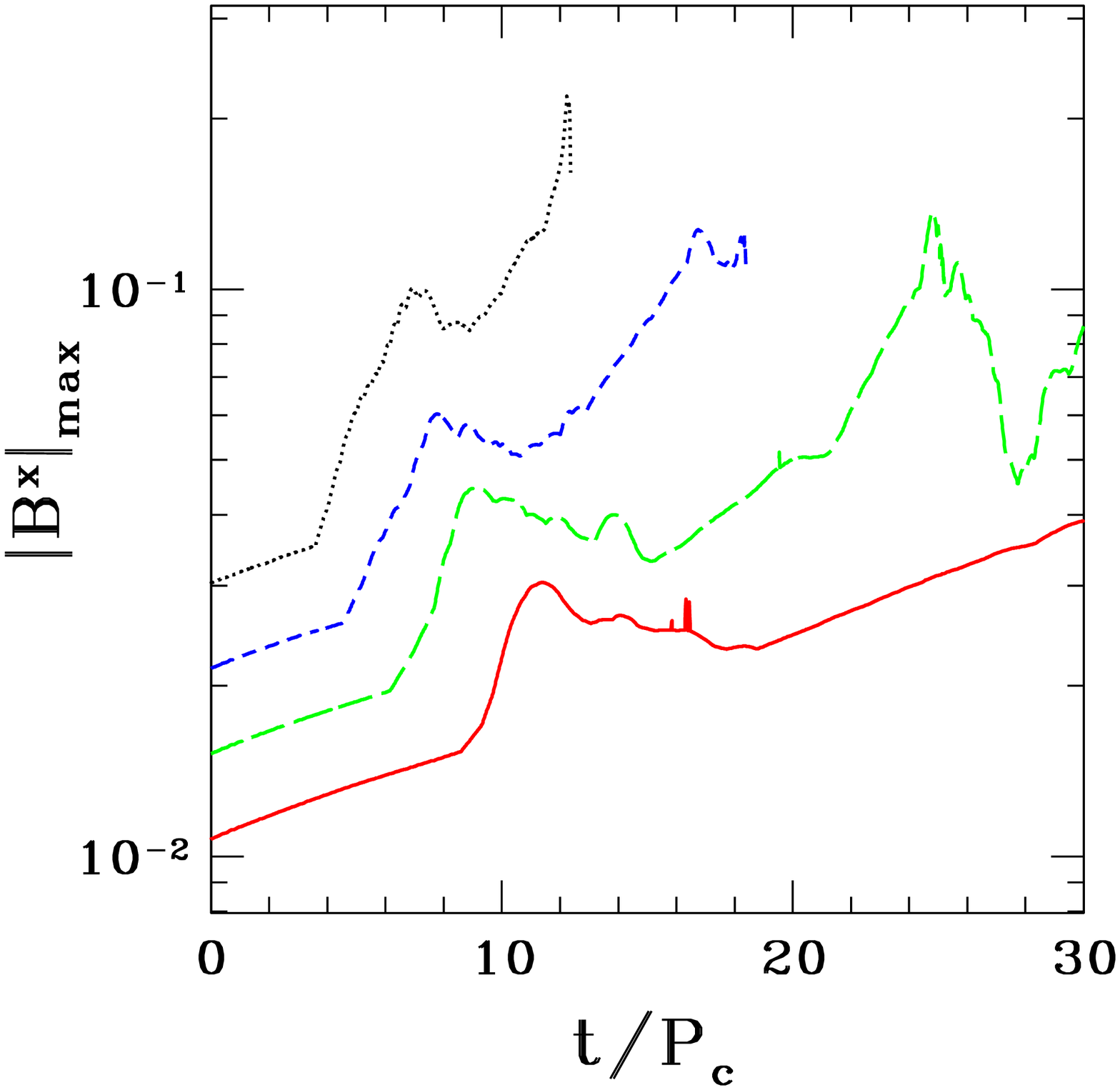}
\end{center}
\caption{{\it Left:}
Selected parameters plotted against scaled time ($t/t_{\rm A}$)
for evolutions of star~A with four different magnetic field strengths:
$C = 1.25 \times 10^{-3}$ (solid lines), $C = 2.5 \times 10^{-3}$ 
(long-dashed lines), $C = 5.0 \times 10^{-3}$ (short-dashed
lines), and $C = 10^{-2}$ (dotted lines).  All runs were performed
with the same resolution ($400^2$ zones with outer boundaries at $20 M$).
When plotted against scaled time, the curves line up at early times 
($t \alt 0.5 t_{\rm A} = 11 P_c$) when the evolution is dominated by 
magnetic winding.  {\it Right:}
Maximum value of $|B^x|$ plotted vs.\ $t/P_c$
for evolutions of star~A with four different magnetic field strengths.  
The line styles are the same as in the left panel. The behavior
of $|B^x|_{\rm max}$ is dominated by the effects of the MRI and thus does 
not scale with the Alfv\'en time. The curves corresponding to the two highest 
values of $C$ (dotted and dashed) terminate at the time when the star 
collapses
\label{scaling}}
\end{figure*}

\subsection{Star~A, comparison of different magnetic field strengths}
\label{scaling_sec}
In order to test the scaling of our results for different values of the
initial magnetic field strength, we have examined three other values of $C$ in
addition to the value of $2.5\times 10^{-3}$ chosen for the results of 
Section~\ref{starA}.  
Namely, we consider $C = \left\{1.25, 2.5, 5.0,  10\right\} \times 10^{-3}$, 
and the results are shown in Fig.~\ref{scaling}.  For the 
portion of the simulations in which magnetic winding dominates, 
the behavior is expected to scale with the Alfv\'en time~\cite{Shapiro}.  
In other words, 
the same profiles should be seen for the same value of $t/t_A$.  From 
the left panel of Fig.~\ref{scaling}, it is evident that
this scaling holds very well for the toroidal field and for the central 
density and lapse, while $t \alt 0.4 t_A$.  
The later evolution is driven mainly by the MRI, which does not scale with the
Alfv\'en time.  The scaling also does not hold during the collapse phase, 
when the evolution is no longer quasi-stationary.  Though the scaling
breaks down at late times in these simulations, the qualitative outcome
is the same in all cases.  
The behavior of $|B^x|_{\rm max}$ for these four different values of $C$ 
is shown in the right panel of Fig.~\ref{scaling}.  The sudden sharp rise 
of $|B^x|_{\rm max}$ 
signals the onset of the MRI, and the approximate agreement of the slopes 
for different values of $C$
indicates that the exponential growth rate of the MRI does not depend on the 
initial magnetic field strength (as expected from the linear analysis).  

\subsection{Star B1}
\label{starB1}

Here, we present results for the evolution of star~B1 with 
$C = 2.5 \times 10^{-3}$.  This run was performed with resolution $400^2$ 
and outer boundaries located at $4.5 R_{\rm eq}$ ($36.4M$).  Since this model is 
not hypermassive, the redistribution of angular momentum through MHD 
effects will not lead to collapse.  However, since this star is ultraspinning 
and angular momentum is conserved in axisymmetric spacetimes, 
it cannot relax to a uniform rotation state everywhere unless a 
significant amount of angular momentum can be dumped to the magnetic 
field.  We find that this model simply seeks out a magnetized equilibrium 
state which consists of a fairly uniformly rotating core surrounded by a 
differentially rotating torus. 

\begin{figure}
\begin{center}
\epsfxsize=2.5in
\leavevmode
\epsffile{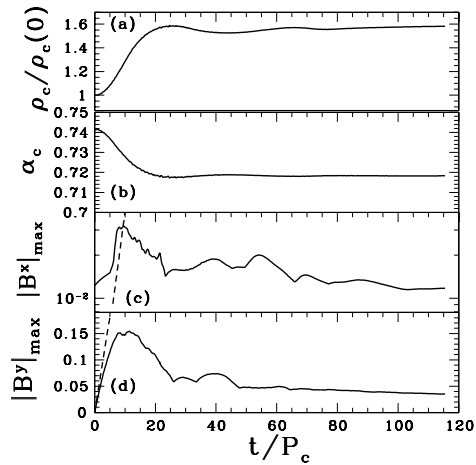}
\caption{Evolution of central rest-mass density $\rho_c$, central lapse 
$\alpha_c$, maximum values of $|B^x|$ and $|B^y|$ for star~B1. The magnetic
fields $|B^x|_{\rm max}$ and $|B^y|_{\rm max}$ are plotted in units of
$\sqrt{\rho_{\rm max}(0)}$.
Note that the lines become fairly horizontal
at late times, indicating that an equilibrium has been reached.
The dashed line in (c) represents an approximate slope of 
$\omega=(0.37/P_c)$ for the exponential growth rate of the MRI, 
$\delta B^x \propto e^{\omega t}$. The dashed line in (d)
represents the predicted growth of $|B^y|_{\rm max}$ from linear theory.
\label{sBfigs}}
\end{center}
\end{figure}

Figure~\ref{sBfigs} presents the evolution of some relevant 
quantities for this case.  From the central density and lapse, it is evident
that the star has settled into a more compact equilibrium configuration.  This
is consistent with the expectation that magnetic braking should transfer 
angular momentum from the core to the outer layers.  A brief episode of 
poloidal magnetic field growth due to the MRI is indicated by the plot of 
$|B^x|_{\rm max}$ in Figure~\ref{sBfigs}.  The instability saturates and 
quickly dies away, leaving the strength of the poloidal field 
largely unchanged.
Early in the evolution, the maximum value of the toroidal component $|B^y|$ 
rises due to magnetic winding.  This growth saturates at 
$t \sim 10 P_c \sim 0.5 t_A$ .  We note, however, that the toroidal magnetic 
field is non-zero in the final equilibrium state, though it is no longer 
growing due to magnetic winding.  

\begin{figure*}
\begin{center}
\epsfxsize=1.8in
\leavevmode
\hspace{-0.7cm}\epsffile{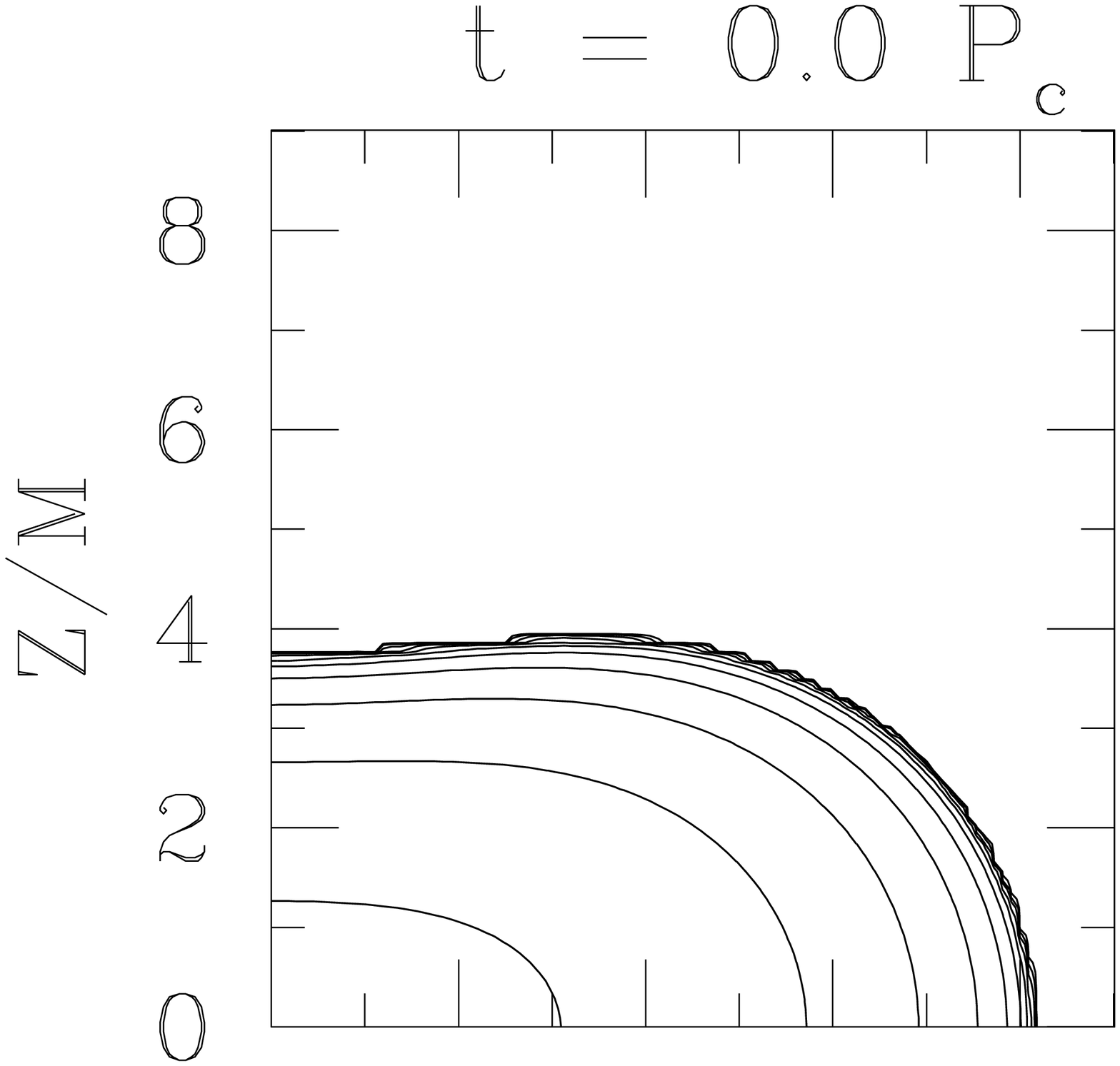}
\epsfxsize=1.8in
\leavevmode
\hspace{-0.5cm}\epsffile{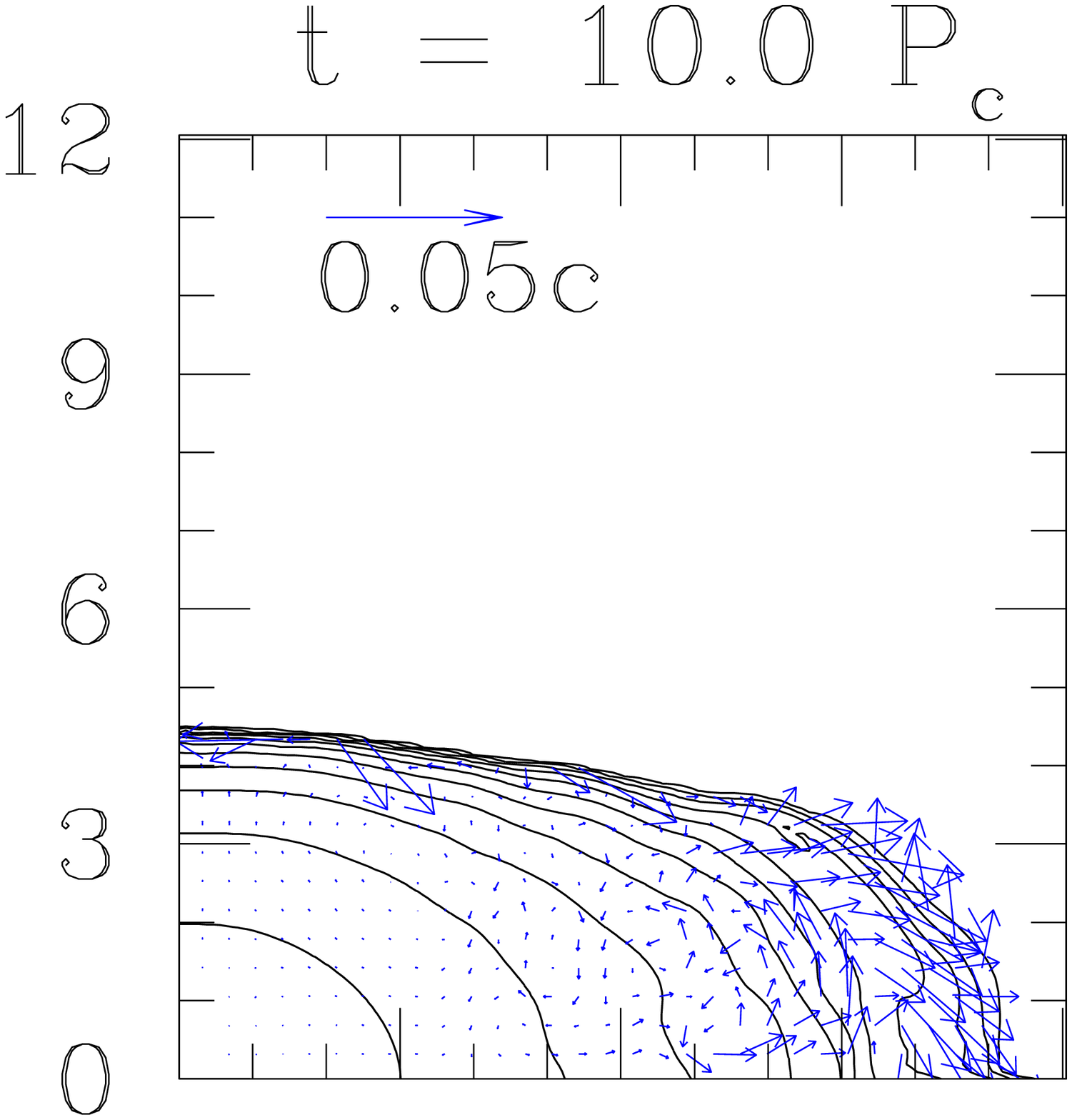}
\epsfxsize=1.8in
\leavevmode
\hspace{-0.5cm}\epsffile{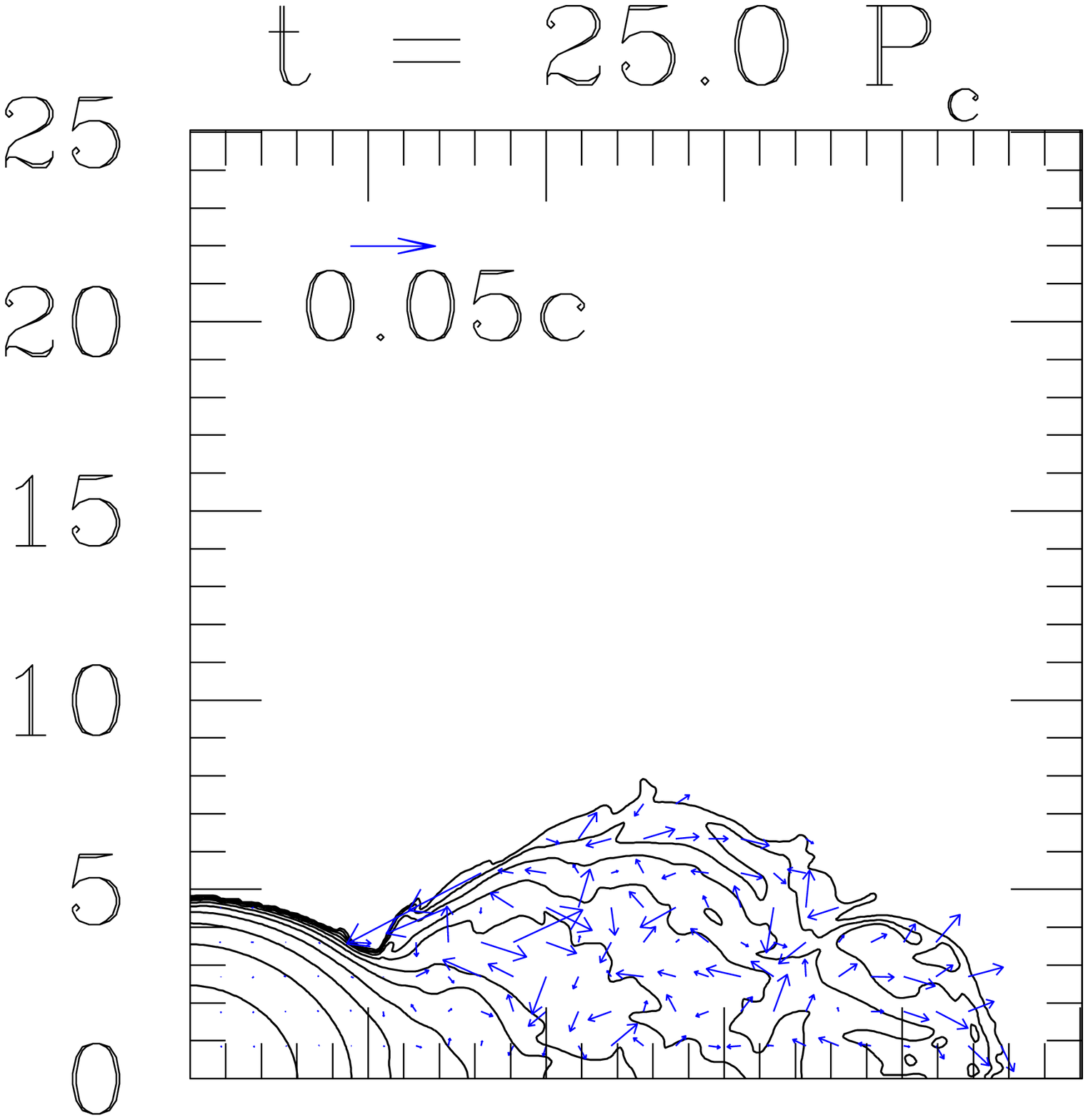} \\
\vspace{-0.5cm}
\epsfxsize=1.8in
\leavevmode
\hspace{-0.7cm}\epsffile{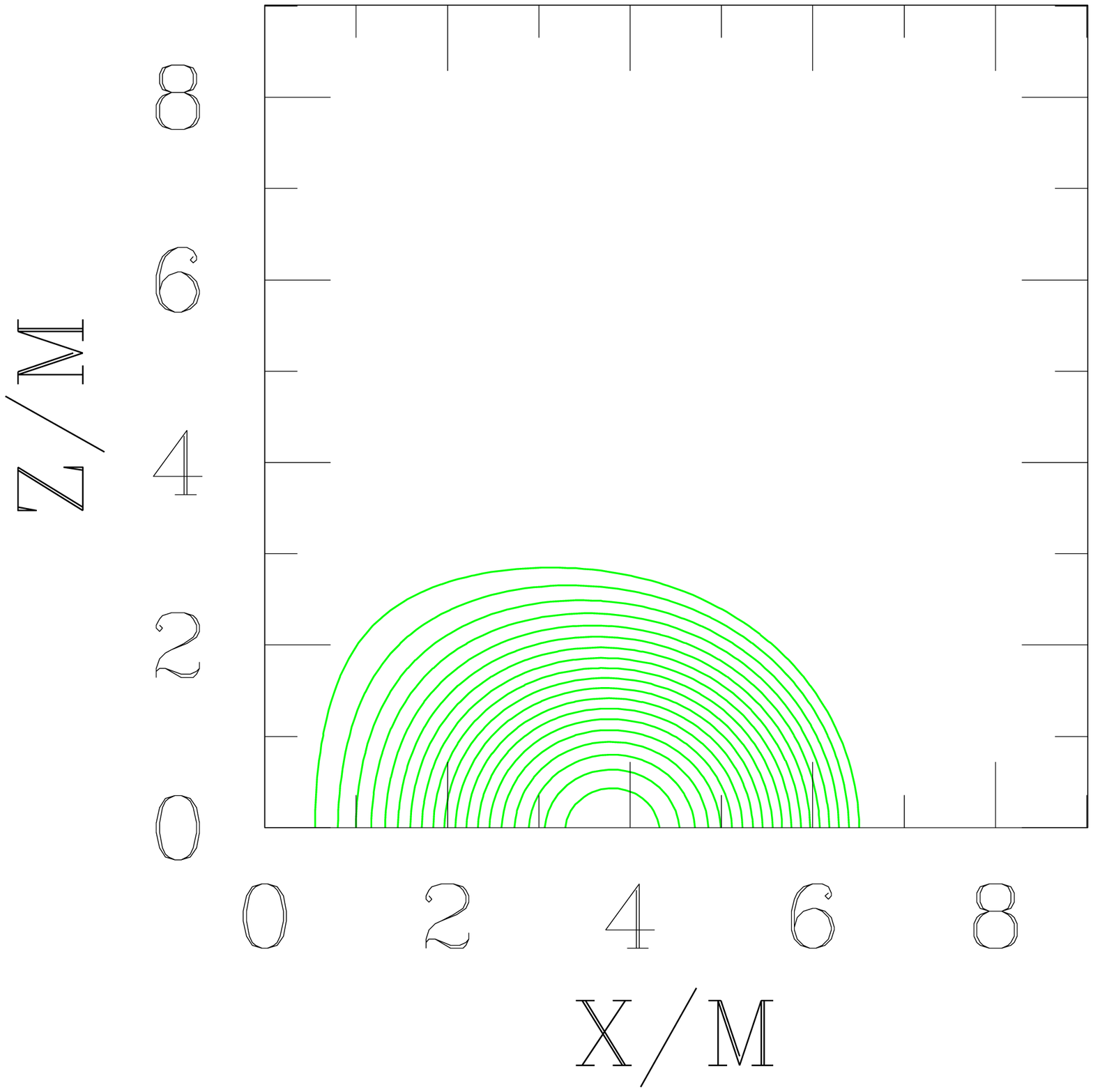}
\epsfxsize=1.8in
\leavevmode
\hspace{-0.5cm}\epsffile{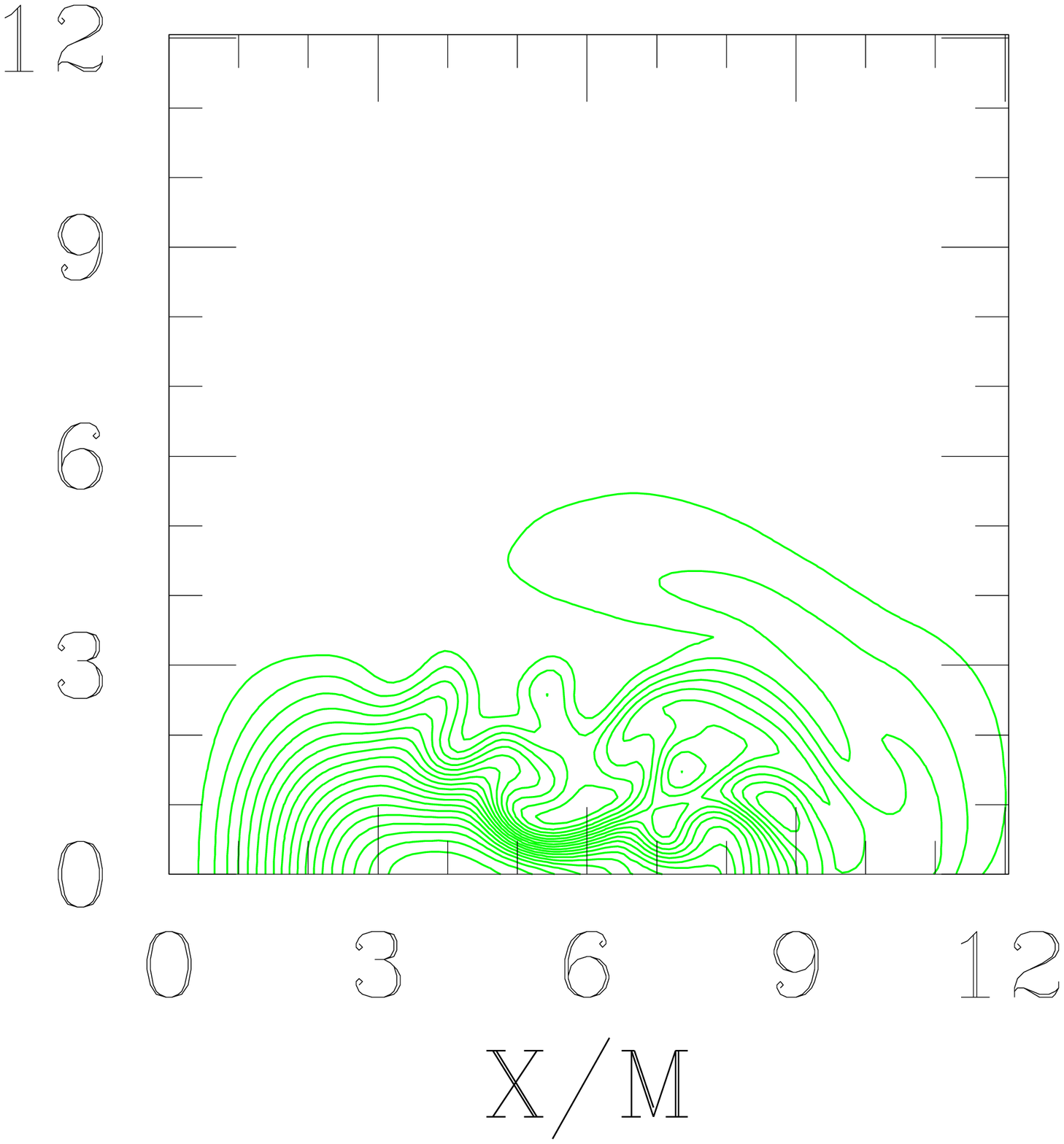}
\epsfxsize=1.8in
\leavevmode
\hspace{-0.5cm}\epsffile{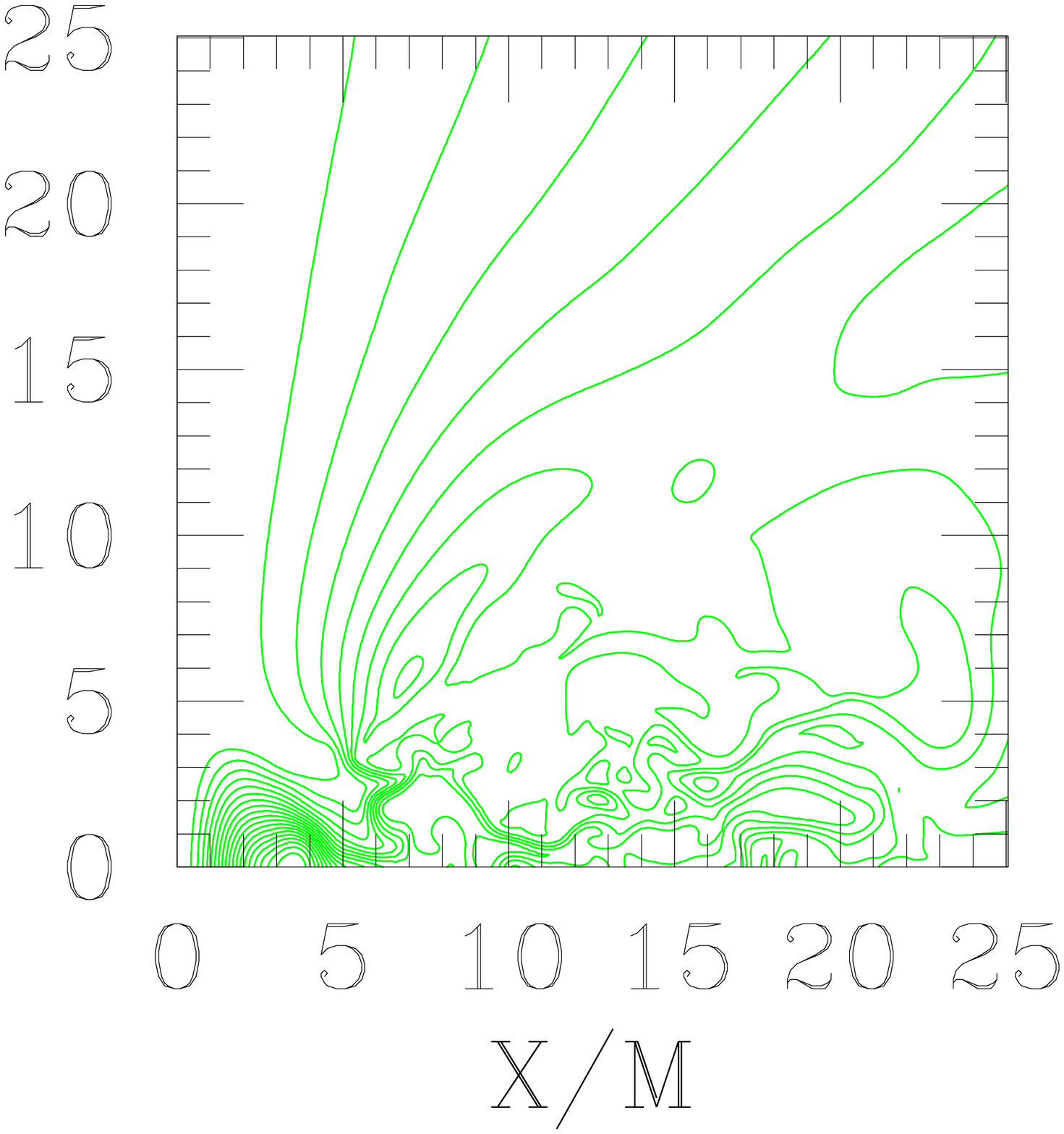}
\end{center}

\begin{center}
\epsfxsize=1.8in
\leavevmode
\hspace{-0.7cm}\epsffile{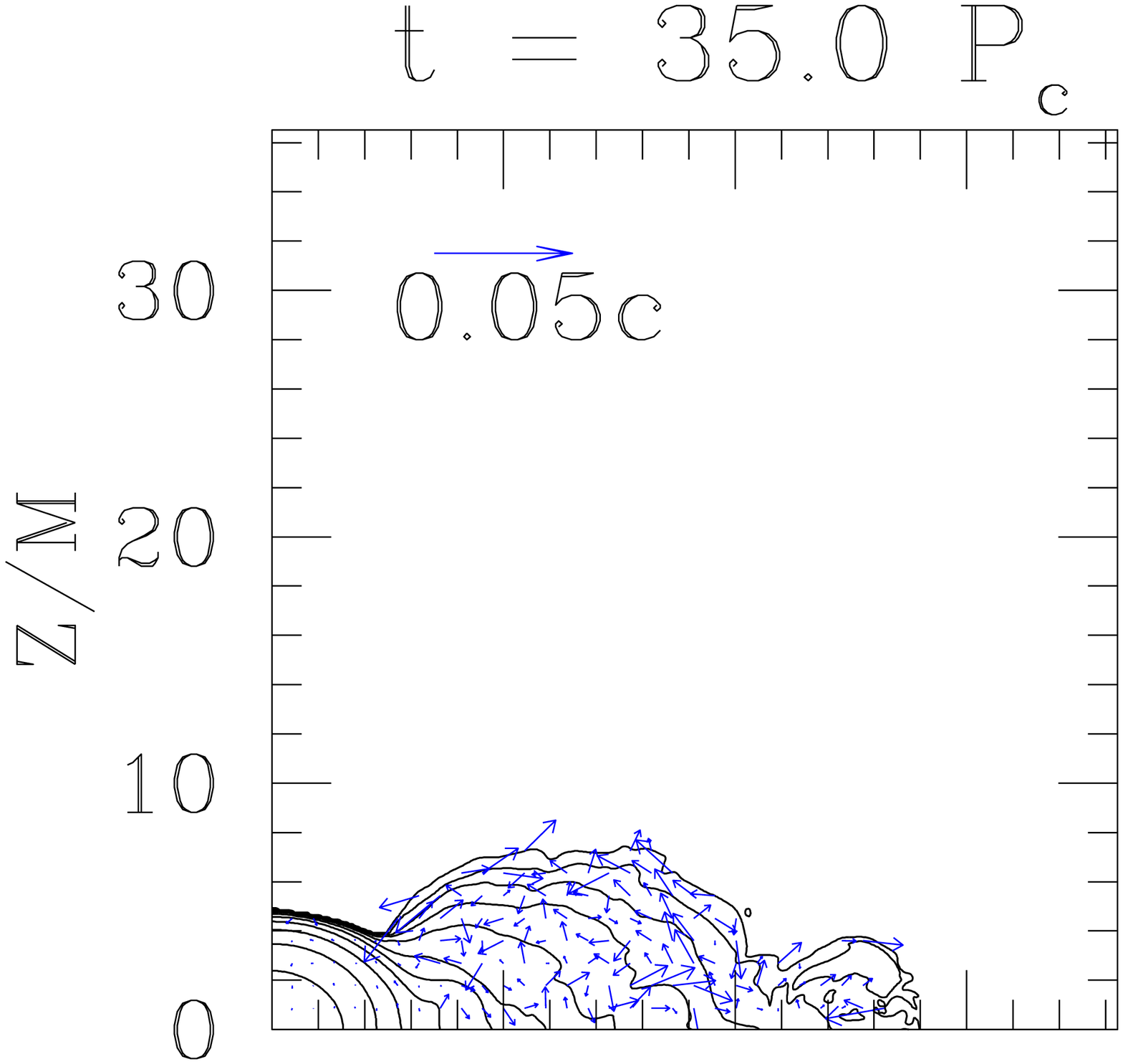}
\epsfxsize=1.8in
\leavevmode
\hspace{-0.5cm}\epsffile{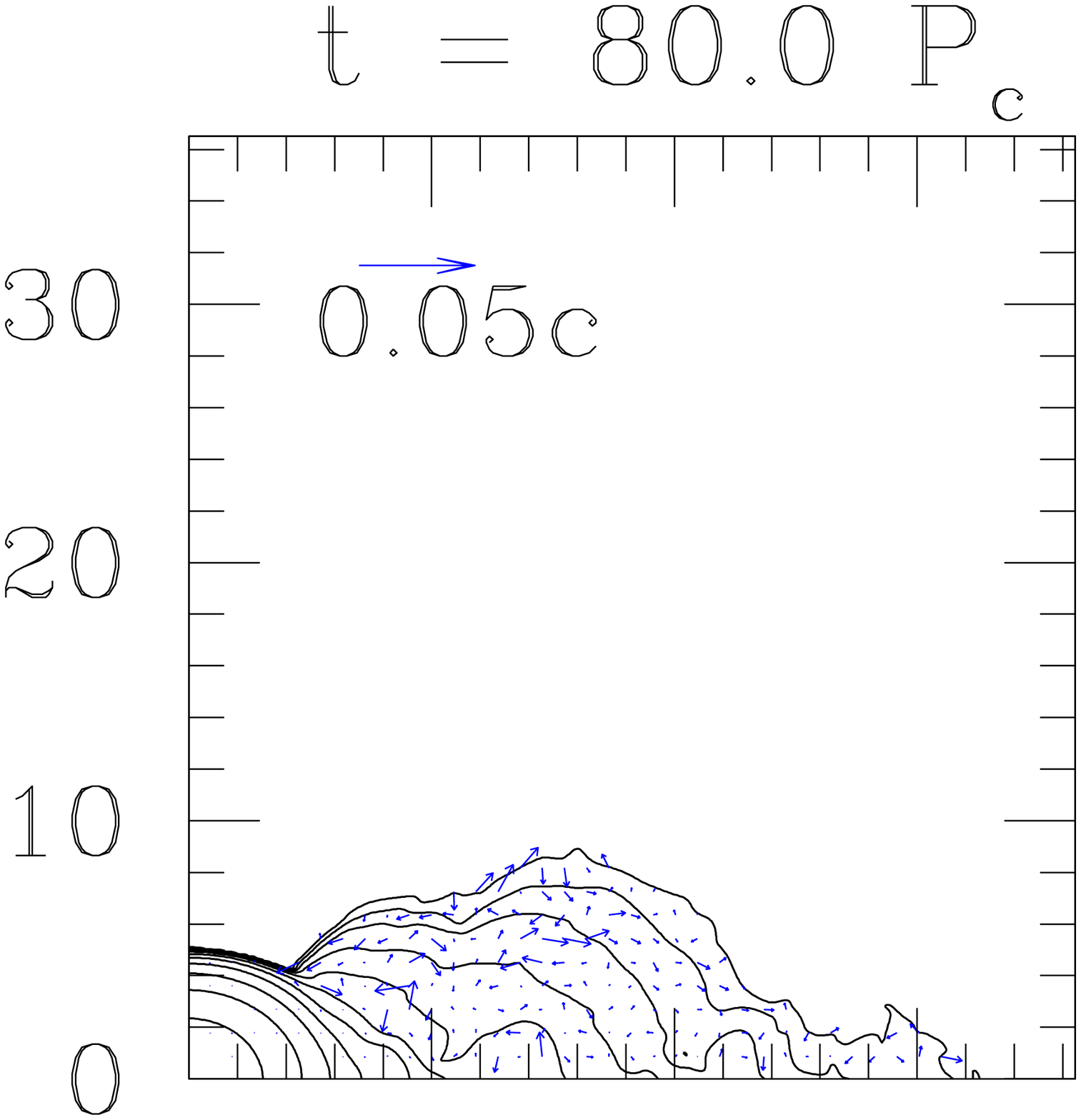}
\epsfxsize=1.8in
\leavevmode
\hspace{-0.5cm}\epsffile{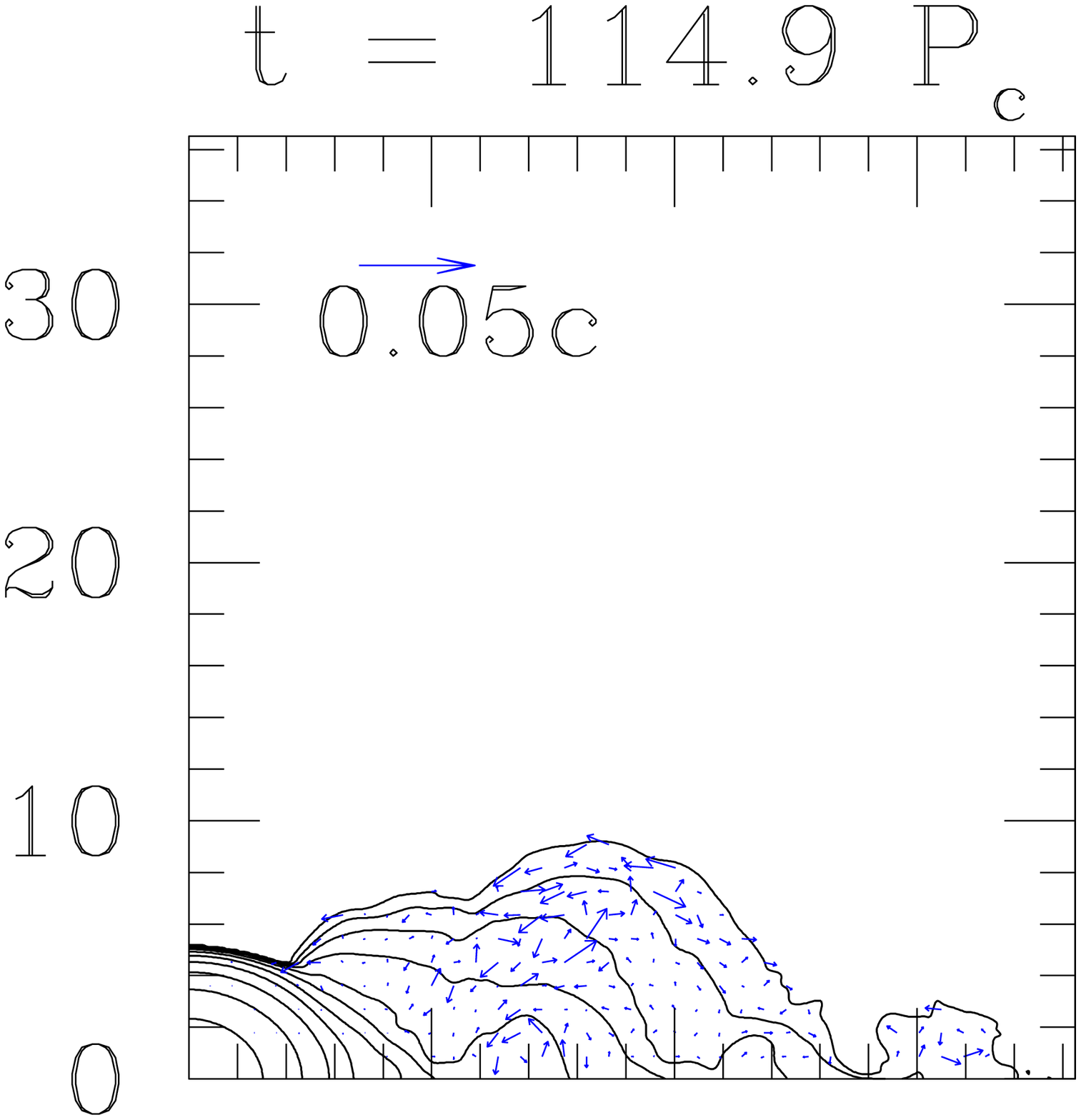} \\
\vspace{-0.5cm}
\epsfxsize=1.8in
\leavevmode
\hspace{-0.7cm}\epsffile{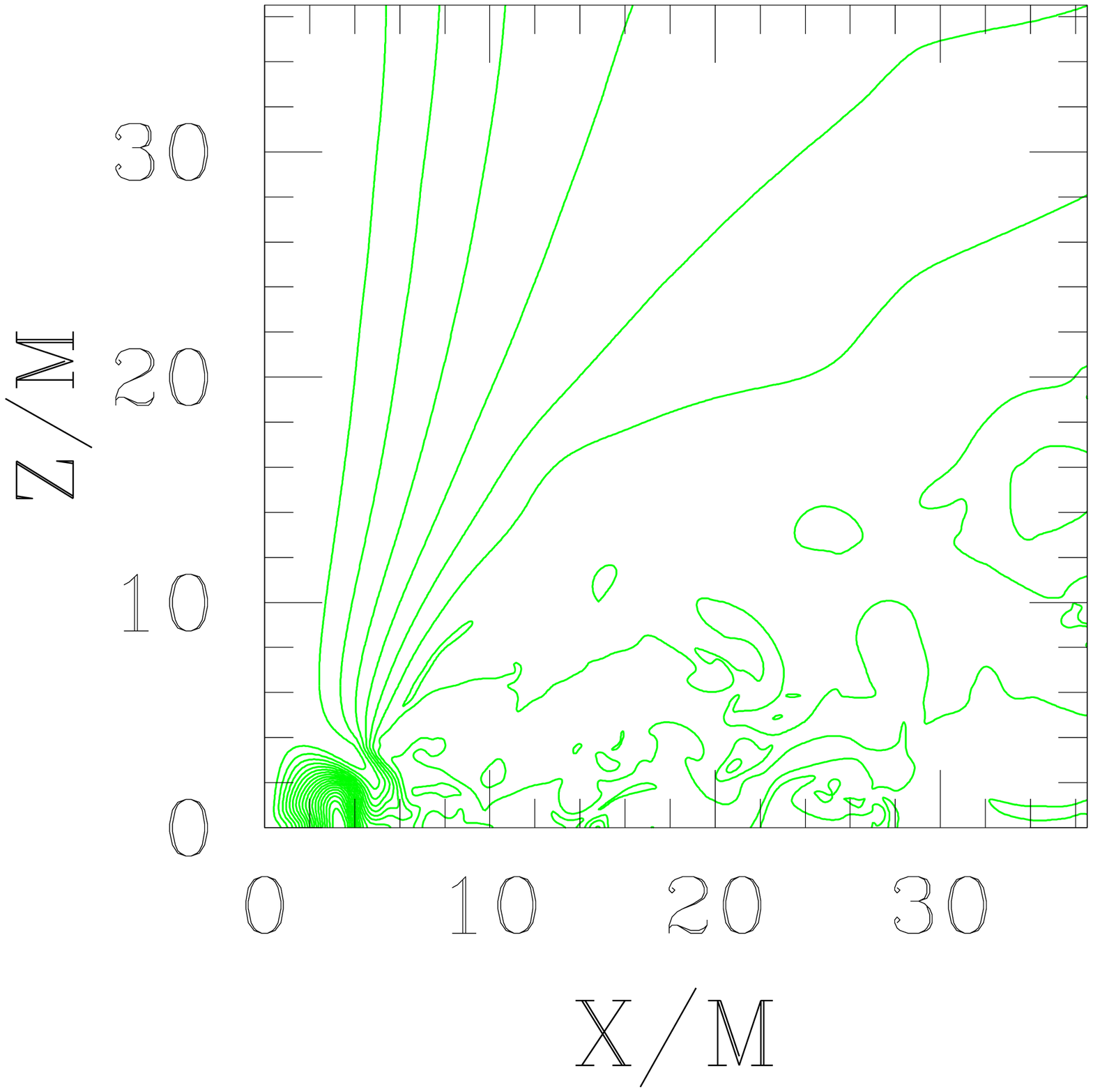}
\epsfxsize=1.8in
\leavevmode
\hspace{-0.5cm}\epsffile{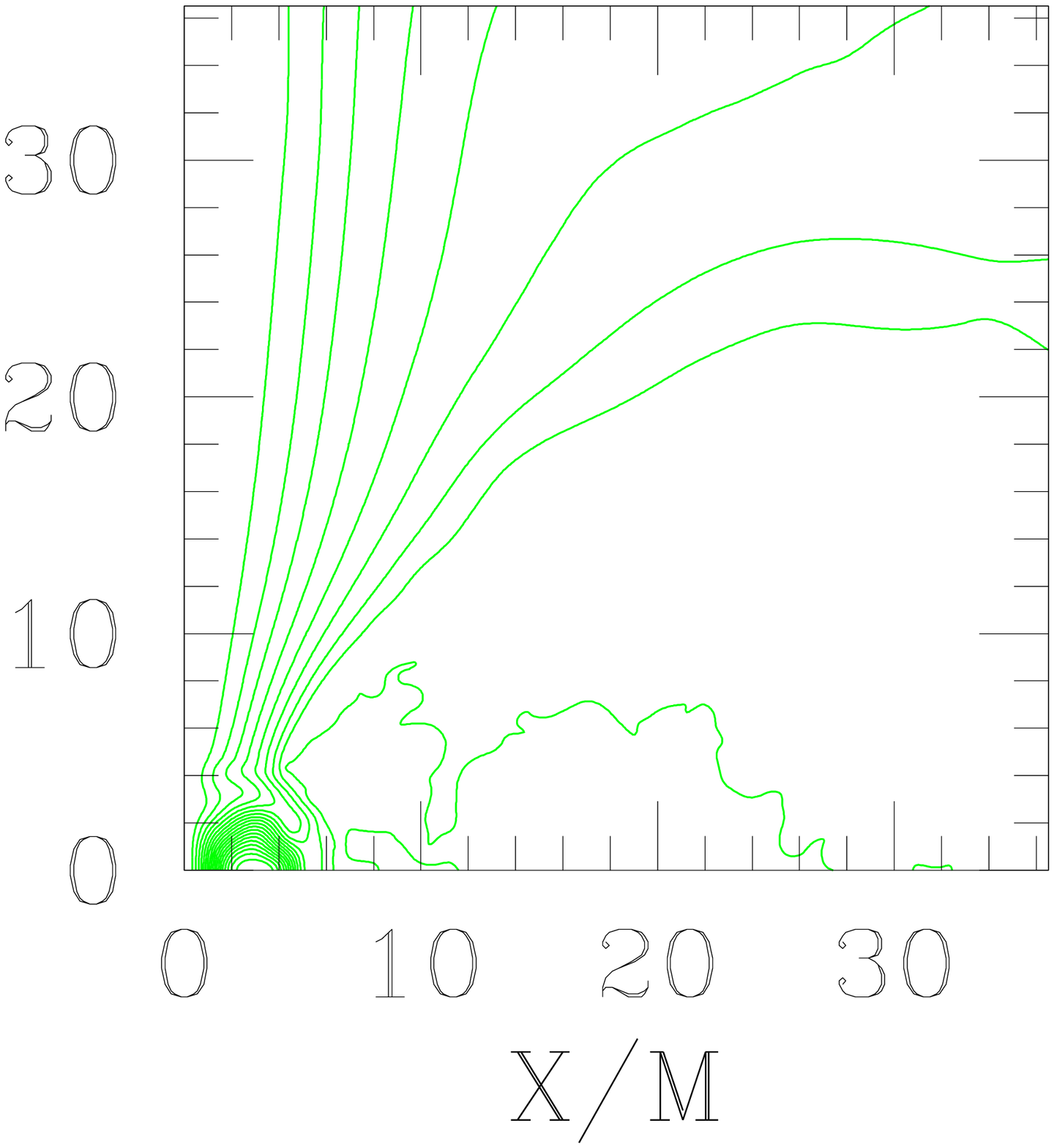}
\epsfxsize=1.8in
\leavevmode
\hspace{-0.5cm}\epsffile{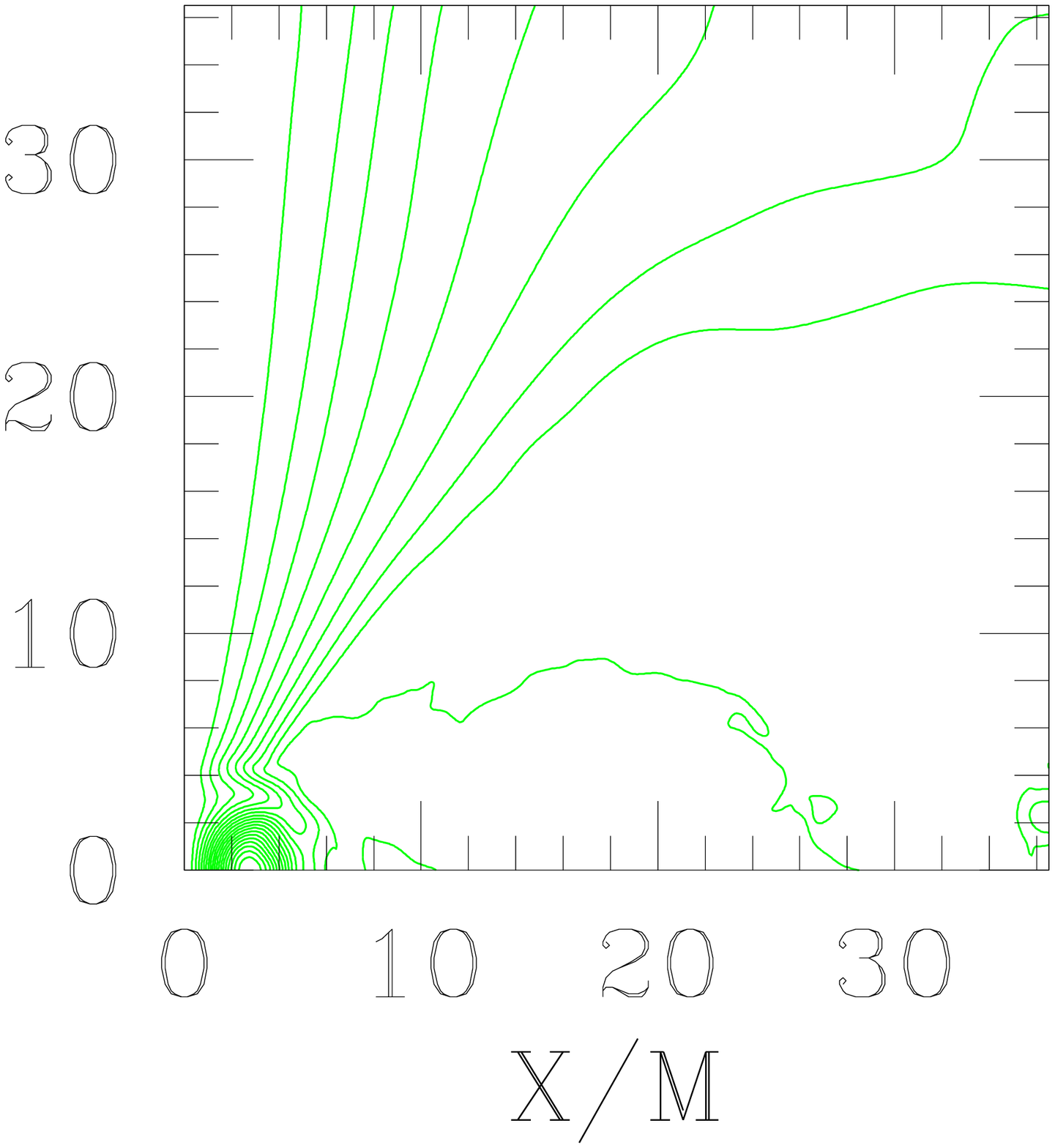}
\caption{Snapshots of density contours and poloidal magnetic field lines for 
star~B1. The first and third rows show snapshots of the rest-mass density
contours and velocity vectors on the meridional plane. The second and fourth
rows show the field lines (lines of constant $A_{\varphi}$)
for the poloidal magnetic field at the same times
as the first and third rows.
The density contours are drawn for $\rho_0/\rho_{\rm max}(0)=
10^{-0.36 i - 0.09}~(i=0$--10).
The field lines are drawn for $A_{\varphi} = A_{\varphi,\rm min}
+ (A_{\varphi,\rm max} - A_{\varphi,\rm min}) i/20~(i=1$--19),
where $A_{\varphi,\rm max}$ and $A_{\varphi,\rm min}$ are the maximum
and minimum values of $A_{\varphi}$ respectively at the given time.
Note that the field lines and the density contours show
little change for $t \agt 35P_c$, indicating that the
star has settled down to an equilibrium state. 
\label{sBmerid}}
\end{center}
\end{figure*}

\begin{figure}
\begin{center}
\epsfxsize=2.5in
\leavevmode
\epsffile{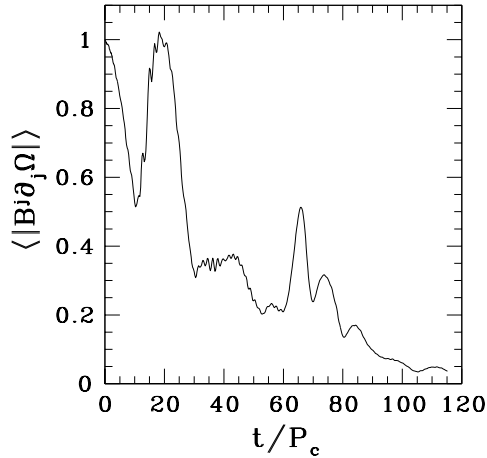}
\caption{Evolution of $\langle |B^j \partial_j \Omega| \rangle$ (normalized 
to unity at $t=0$). Note that this quantity drops toward zero at late time, 
indicating that the star is driven to a differentially rotating equilibrium 
state in which 
$\Omega$ is constant along the magnetic field lines. \label{bdotgradOm}}
\end{center}
\end{figure}

Snapshots of the evolution in the $x$-$z$ plane are shown in 
Fig.~\ref{sBmerid}. The density contours for times $t=0$ through
$25.0 P_c$ show that angular momentum redistribution leads to the formation 
of a more compact star surrounded by a torus.  At $t=10 P_c$, the distortions 
of the magnetic field lines due to the MRI are clearly visible.  As the disk 
expands, magnetic field lines attached to this low density material open 
outward, eventually leading to the field structure seen in the last three
times shown in Fig.~\ref{sBmerid}, for which some field lines are still 
confined inside the star while others have become somewhat collimated 
along the $z$-axis.   For $t \agt 35 P_c$, the density contours and poloidal
magnetic field lines change very little, indicating that the system has
reached an equilibrium state which is quite different from the initial state.
We note that a significant toroidal 
field persists at late times when the system has essentially settled down to 
a final state.  In addition, the final configuration also has significant
differential rotation, especially in the outer layers.
However, this differential 
rotation no longer winds up the magnetic field lines (i.e., the 
toroidal field strength does not grow).  This is because the rotation 
profile has adjusted so that $\Omega$ is approximately constant along 
magnetic field lines.  This is demonstrated in Fig.~\ref{bdotgradOm}, 
which shows that the density-weighted average of the quantity 
$B^j \partial_j \Omega$ approaches zero
at late times.  Since the rotation 
profile is adapted to the magnetic field structure, a stationary final 
state is reached which allows differential rotation and a nonzero 
toroidal field.  

%In cgs units, we find that for the initial field 
%considered here, the final field strength is
%\begin{equation}
%  |B_{\rm final}| \sim 10^{17} \left(\frac{2 M_{\odot}}{M}\right)~{\rm G} \ .
%\end{equation}
%This field is comparable to the field strength of a magnetar. 
%Since the strength 
%of the initial seed magnetic field is much smaller than the strength when 
%it saturates, it is possible that the final equilibrium state will be the same 
%even if the initial seed field is much smaller than the present value. 
%If this is true, a new-born neutron star with mass and angular momentum 
%distribution similar to star~B1 is likely to end up as a magnetar due 
%to MHD processes.

\subsection{Star B2}
\label{sec:starB2}

\begin{figure}
\vspace{-4mm}
\begin{center}
\epsfxsize=1.8in
\leavevmode
\hspace{-0.7cm}\epsffile{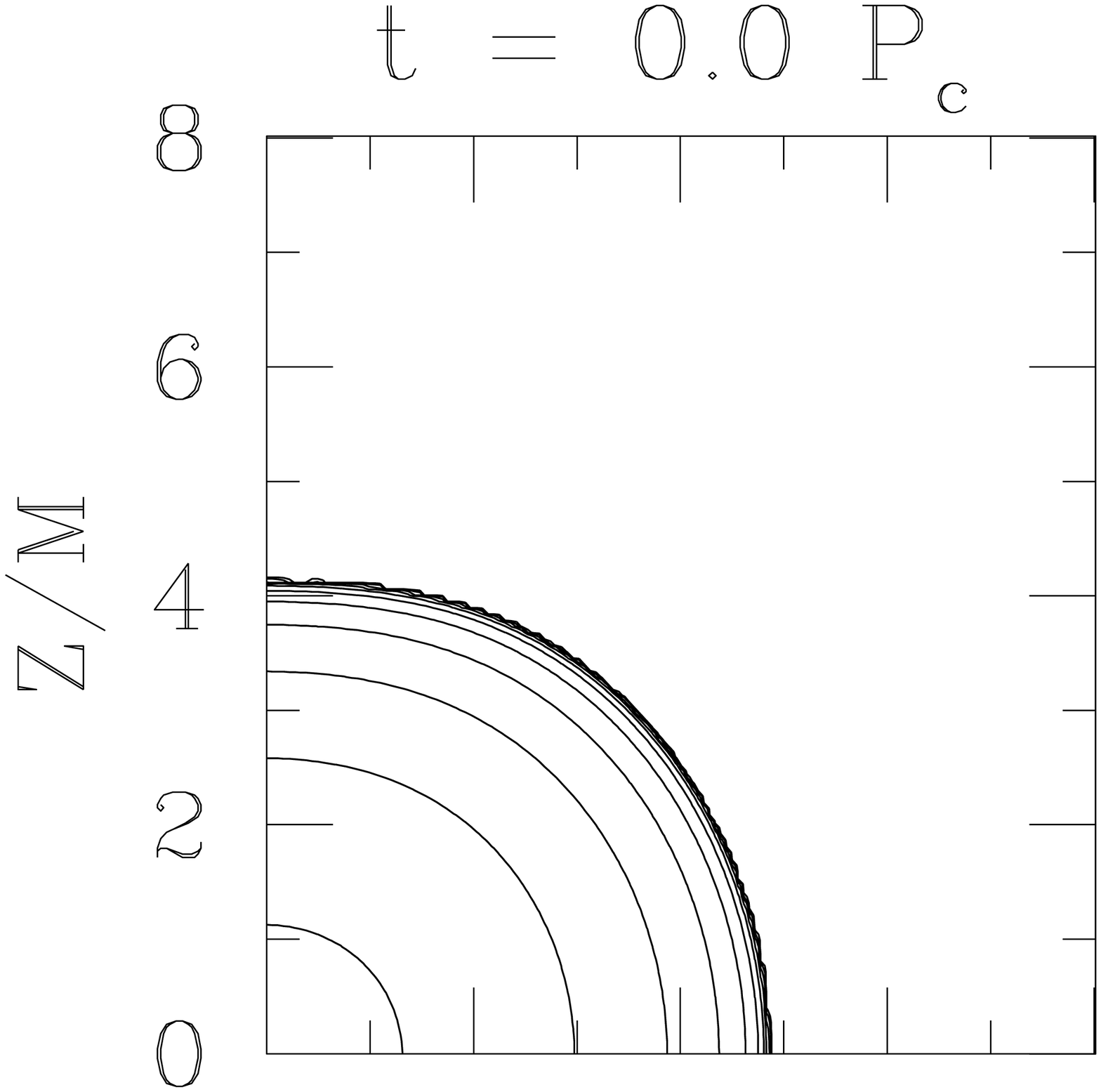}
\epsfxsize=1.8in
\leavevmode
\hspace{-0.5cm}\epsffile{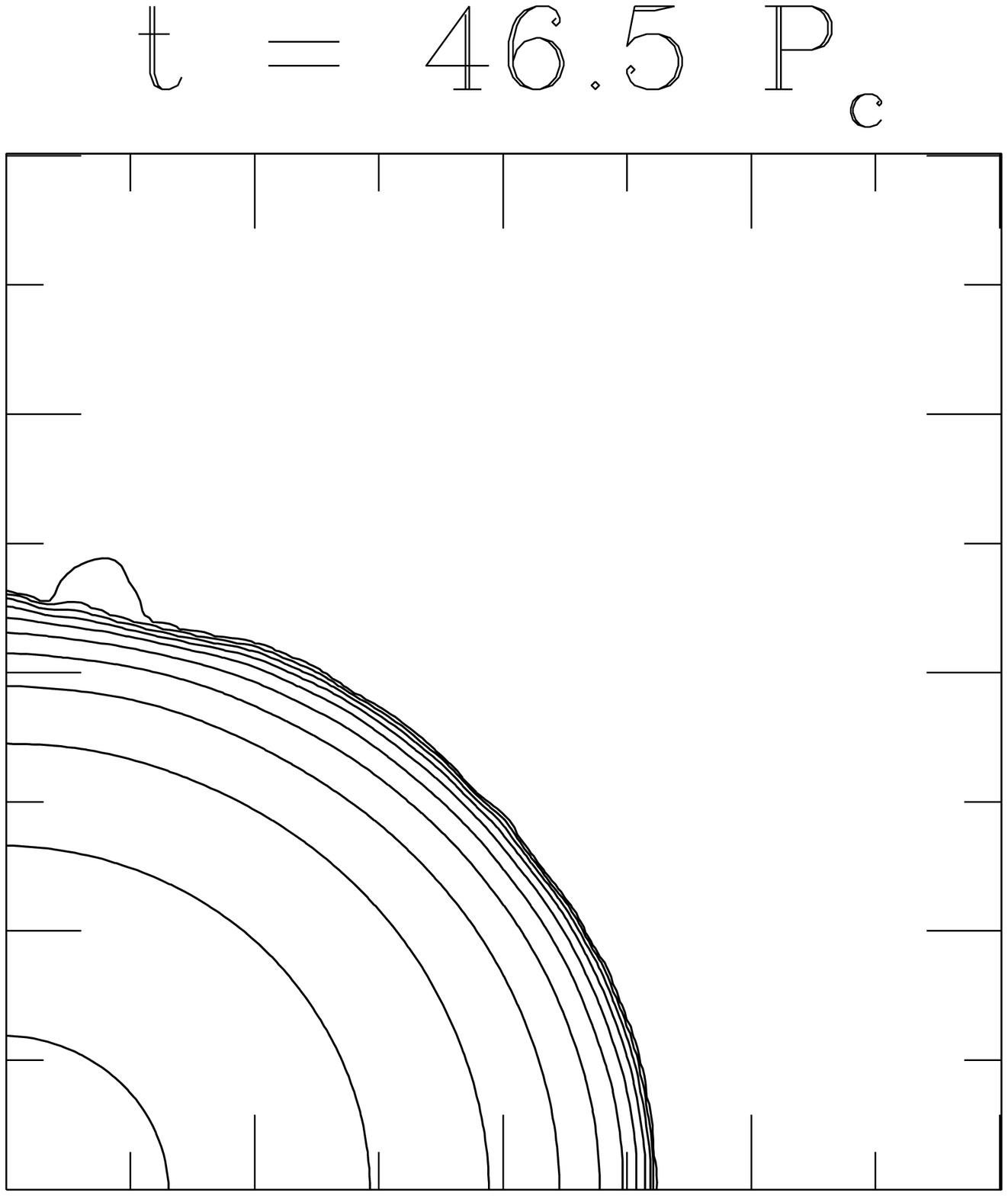} \\
\vspace{-0.3cm}
\epsfxsize=1.8in
\leavevmode
\hspace{-0.7cm}\epsffile{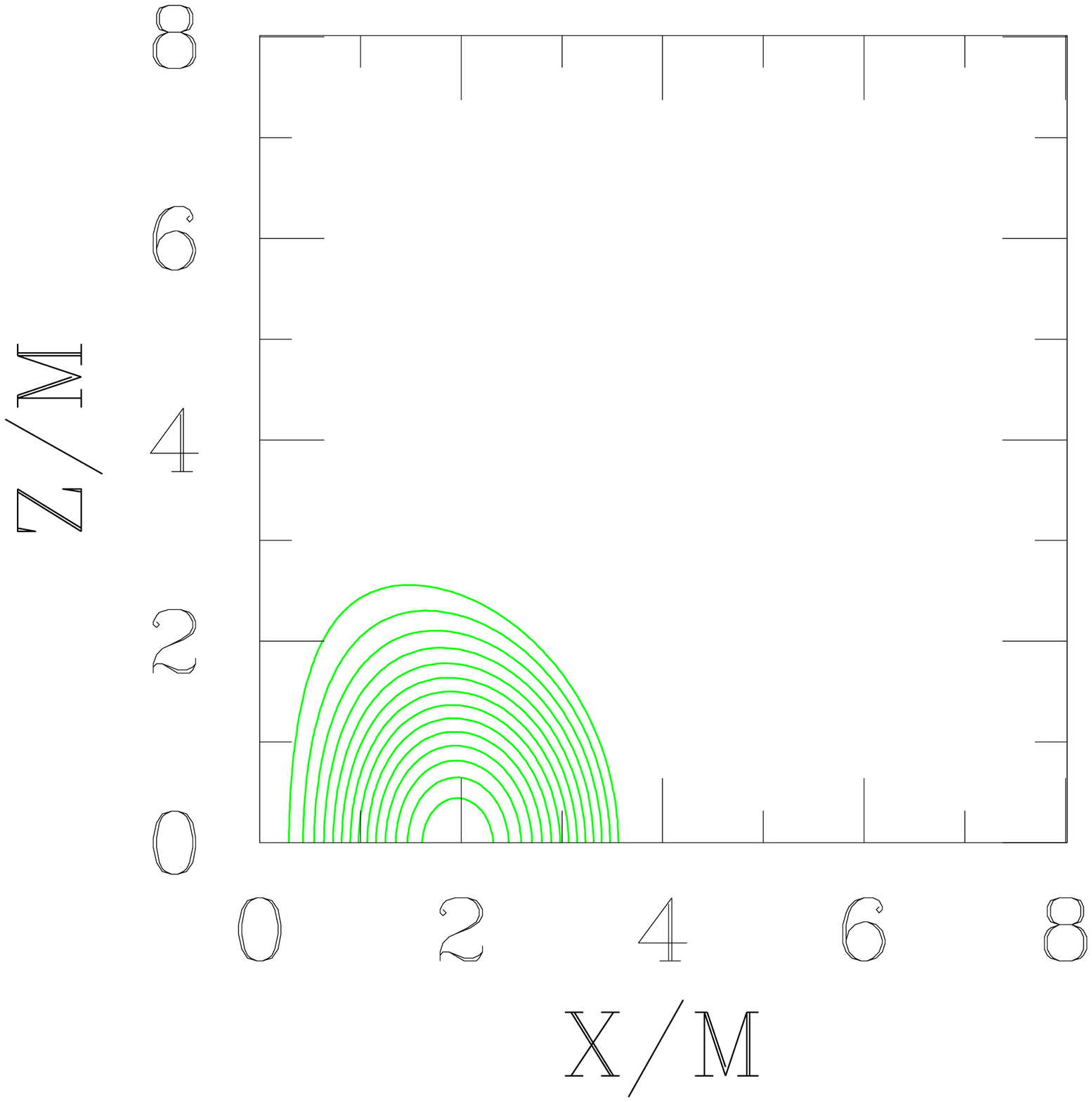}
\epsfxsize=1.8in
\leavevmode
\hspace{-0.5cm}\epsffile{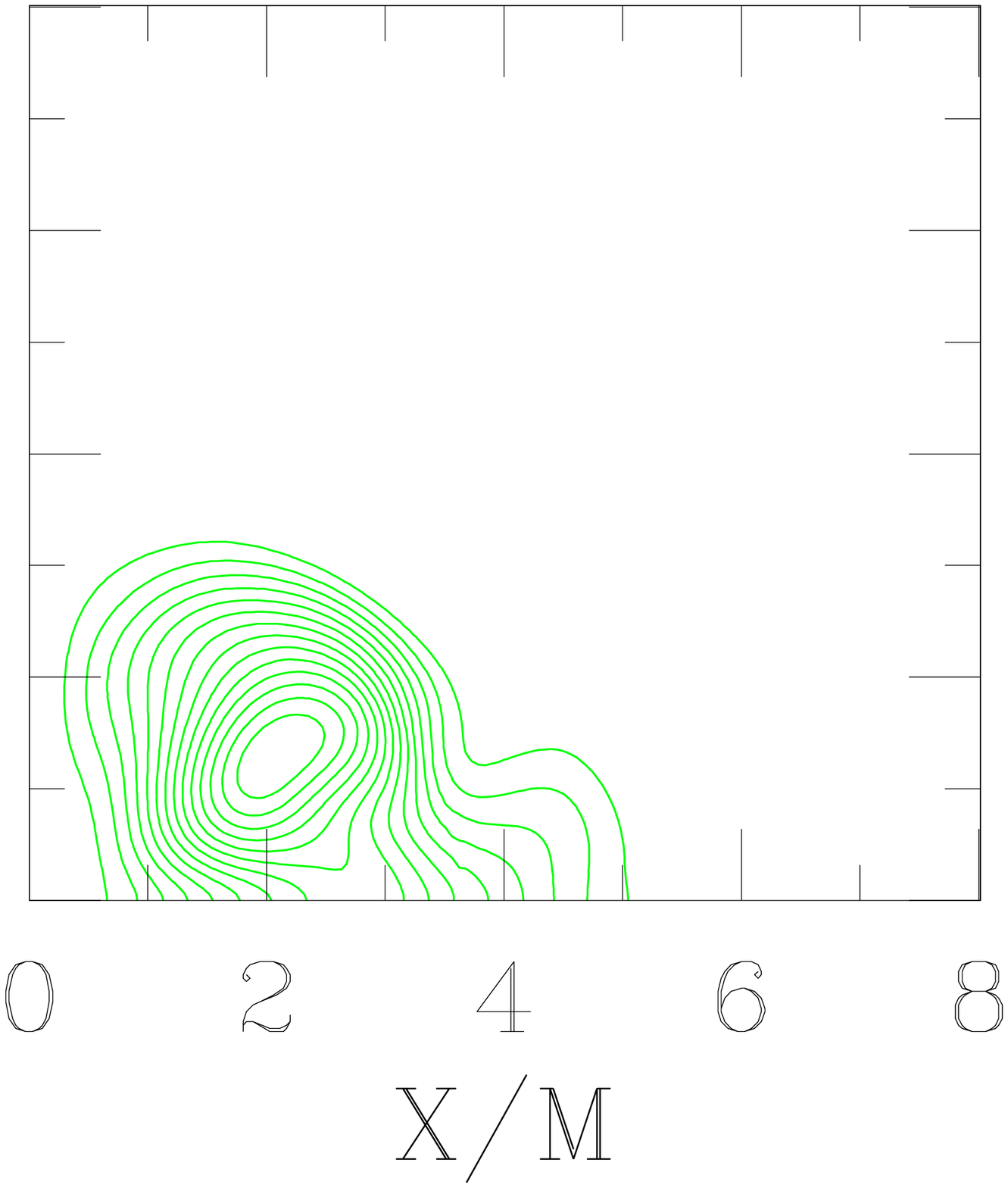}
\caption{Snapshots of the rest-mass density contours and poloidal
magnetic field lines for star~B2 at times $t=0$ and $t=46.5P_c$. 
The first row shows snapshots of the rest-mass density
contours on the meridional plane. The second 
row shows the corresponding field lines 
for the poloidal magnetic field at the same times.
The density contours are drawn for $\rho_0/\rho_{\rm max}(0)=
10^{-0.36 i - 0.09}~(i=0$--10), where $\rho_{\rm max}(0)$ is the
maximum rest-mass density at $t=0$.
The field lines are drawn for $A_{\varphi} = A_{\varphi,\rm min}
+ (A_{\varphi,\rm max} - A_{\varphi,\rm min}) i/15~(i=1$--14),
where $A_{\varphi,\rm max}$ and $A_{\varphi,\rm min}$ are the maximum
and minimum values of $A_{\varphi}$, respectively, at the given time. 
The meridional components of the velocity (which are zero initially) 
at $t=46.5P_c$ are very small and so are not shown here.
\label{fig:cont_b2}}
\end{center}
\end{figure}

Both stars~B1 and B2 are nonhypermassive. However, star~B1 is 
ultraspinning, whereas B2 is normal. 
We evolve B2 with a seed magnetic field strength $C=2.5\times 
10^{-3}$. Our simulation shows that this star evolves 
to a uniformly rotating configuration with little structural 
change.

Figure~\ref{fig:cont_b2} shows the density contours and poloidal magnetic 
field at the initial time ($t=0$) and at $t=46.5P_c\approx 5.8t_A$. 
We see that the density profile of the star does not change 
appreciably. This is not surprising since the main effect of 
the MHD processes is to redistribute the angular momentum 
inside the star. However, the rotational kinetic energy of 
star~B2 is not very large (the initial $T/|W|=0.040$).
Hence, the change of the centrifugal force inside the star 
as a result of angular momentum transport does not 
disturb the initial equilibrium significantly, unlike the cases of 
stars~A and B1.

\section{Discussion and Conclusions}
\label{sec:conclusions}
We have discussed in detail the evolution of magnetized HMNSs as
first reported in~\cite{DLSSS1,GRB2}.  In addition, we have performed 
simulations of two differentially rotating, but nonhypermassive, neutron 
star models with the same initial magnetic field geometry.  These 
simulations have revealed a rich variety of behavior with possible 
implications for astrophysically interesting systems such as binary 
neutron star remnants, nascent neutron stars, and GRBs, 
where magnetic fields and strong gravity both play important roles.

The hypermassive model considered here, star~A,
collapses to a BH due to the influence of the initial seed magnetic field.  
The early phase of evolution is dominated by 
magnetic winding, which proceeds until the back-reaction
on the fluid becomes strong enough that the growth of the toroidal field 
ceases.  This happens after roughly one Alfv\'en time.  After several 
rotation periods, we also see the effects of the MRI, which 
leads to turbulence.
Eventually, the inner core of star~A becomes unstable and
collapses to a BH.  Surrounding the BH, a significant amount of material 
remains in a magnetized torus which has been heated considerably by 
shocks resulting from the turbulent motions of the fluid.  This final 
state consisting of a BH surrounded by a massive, hot accretion disk may
be capable of producing highly relativistic outflows and 
a fireball (either through
$\nu-\bar{\nu}$ annihilation or MHD processes) and is hence a promising
candidate for the central engine of short-hard GRBs.  This model predicts 
that such GRBs should accompany a burst of gravitational radiation and neutrino 
emission from the HMNS delayed collapse. 

The behavior of the nonhypermassive, ultraspinning star~B1 under the 
influence of a 
seed magnetic field is quite different.  Magnetic braking and the MRI 
operate in this model as well, leading to a mild contraction of the 
inner core and the expansion of the outer layers into a high angular
momentum torus-like structure.  
The final state consists of a fairly uniformly rotating core surrounded
by a differentially rotating torus.  The remaining differential rotation 
does not shear the magnetic field lines, so that the toroidal field 
settles down.  On the other hand, the normal star B2 
simply evolves to a uniformly rotating configuration. 

Two issues in particular warrant further study.  The first is the 
scaling behavior of our solutions.  We begin our simulations with a 
seed magnetic field which, though far too weak to be dynamically 
important, may be significantly larger than magnetic fields present 
in HMNSs formed through stellar collapse or a binary neutron star merger.
We have demonstrated that, by varying the strength of the initial
magnetic field through a factor of $\sim 3$ (See Fig.~\ref{scaling}), 
our evolution obeys the expected scaling during the magnetic winding 
phase, and the qualitative outcome of the simulations remains the same.  
However, since the MRI grows on a timescale $\sim \mbox{few}\times P_c$ 
regardless of the initial magnetic field strength, it is possible that, 
for very weak initial fields, the effects of the MRI could dominate the 
evolution long before the effects of magnetic braking become important.
In this case, the scaling of our numerical results with the Alfv\'en
time (relevant for magnetic winding) may break down.  

Another concern is the effect
on our evolutions of relaxing the axisymmetry assumption.  Rapidly
and differentially rotating neutron stars may be subject to bar
and/or one-armed spiral mode instabilities which could affect the 
dynamics (though star~A was shown in~\cite{BSS,DLSS} to be stable 
against such instabilities, at least on dynamical timescales).  
Additionally, the development of the MRI in 2D differs from the 
3D case~\cite{hgb95}.  Turbulence arises and persists more readily 
in 3D due to the lack of symmetry~\cite{moffatt78,hb92}.
However, McKinney and Gammie~\cite{mg04} 
have performed axisymmetric simulations of magnetized tori accreting 
onto Kerr BHs and have found good quantitative agreement with the 3D 
results of De Villiers and Hawley~\cite{dvhkh05} on the global 
features of the evolution.  Though simulations in full 3D will 
eventually be necessary to capture the full behavior of magnetized 
HMNSs, the 2D results presented here likely provide (at least) 
a good qualitative picture.  

\ack
It is a pleasure to thank C.~Gammie for useful suggestions and discussions.
Numerical computations were performed at the National Center for
Supercomputing Applications at the University
of Illinois at Urbana-Champaign (UIUC), and
on the FACOM xVPP5000 machine at
the data analysis center of NAOJ and the NEC SX6 machine in ISAS,
JAXA. This work was in part supported by NSF Grants PHY-0205155
and PHY-0345151, NASA Grants NNG04GK54G and NNG046N90H
at UIUC, and
Japanese Monbukagakusho Grants (Nos.\ 17030004 and 17540232).

\end{document}